\documentclass[]{aa}
\usepackage{graphicx}
\usepackage{eufrak}
\usepackage{txfonts}
\usepackage{natbib}
\bibpunct{(}{)}{;}{a}{}{,}

\begin{document}

\title{Dust-driven mass loss from carbon stars\\ as a function of stellar parameters}
\subtitle{II. Effects of grain size on wind properties}
\titlerunning{Mass Loss from Carbon Stars as Function of Stellar Parameters. II.}

\author{L. Mattsson\inst{1,2}\thanks{\email{mattsson@dark-cosmology.dk}}\and S. H\"ofner\inst{2}}
\institute{DARK Cosmology Centre, Niels Bohr Institute, University of Copenhagen, Juliane Maries Vej 30, DK-2100, Copenhagen \O, Denmark
\and
Dept. Physics and Astronomy, Div. of Astronomy and Space Physics, Uppsala University, Box 516, SE-751 20 Uppsala, Sweden}

\offprints{Lars Mattsson}

\date{Received date; accepted date}

\abstract
{It is well established that the winds of carbon-rich AGB stars
(carbon stars) can be driven by radiation pressure on grains of
amorphous carbon and collisional transfer of momentum to the gas.
This has been demonstrated convincingly by different numerical wind
models that include time-dependent dust formation. To simplify the
treatment of dust opacities, radiative cross sections are usually
computed using the assumption that the dust grains are small
compared to wavelengths around the stellar flux maximum. Considering
the typical grain sizes that result from these models, however, the
applicability of this small-particle limit (SPL) seems
questionable.}
{We explore grain size effects on wind properties of
carbon stars, using a generalized description of radiative cross
sections valid for particles of arbitrary sizes. The purpose of the
study is to investigate under which circumstances the SPL may give
acceptable results, and to quantify the possible errors that may
occur when the SPL does not hold.}
{The time-dependent description of grain growth in our detailed
radiation-hydrodynamical models gives information about dust
particle radii in every layer at every instant of time. Theses grain
radii are used for computing opacities and determining the
radiative acceleration of the dust-gas mixture. From the large
number of models presented in the first paper of this series (based
on SPL dust opacities; Mattsson et al.~2010) we selected two
samples, i.e., a group of models with strong, well-developed
outflows that are probably representative of the majority of
wind-forming models, and another group, close to thresholds in
stellar parameter space for dust-driven winds, which are referred
to as critical cases.}
{We show that in the critical cases the effect of the
generalized description of dust opacities can be significant,
resulting in more intense mass loss and higher wind velocities
compared to models using SPL opacities. For well-developed winds,
however, grain size effects on mass-loss rates and wind velocities
are found to be small. Both groups of models tend towards lower
degrees of dust condensation compared to corresponding SPL models,
owing to a self-regulating feedback between grain growth and radiative
acceleration. Consequently, the "dust-loss rates" are lower in the
models with the generalized treatment of grain opacities.}
{We conclude that our previous results on mass-loss rates obtained
with SPL opacities are reliable within a wide region of stellar
parameter space, except for critical cases close to thresholds of
dust-driven outflows where SPL models will tend to underestimate the
mass loss rates and wind velocities. 
}
\keywords{Stars: AGB and post-AGB -- Stars: atmospheres --
          Stars: carbon -- Stars: circumstellar matter --
          Stars: evolution -- Stars: mass loss --2
          Hydrodynamics -- Radiative transfer}

\maketitle

\section{Introduction}

Winds of carbon stars are usually considered to {be
dust-driven winds}. Stellar photons, incident on dust particles, will
lead to a radiative acceleration of the grains away from the star,
and, subsequently, momentum will be transferred to the surrounding
gas by gas--grain collisions. Pulsation-induced atmospheric shock
waves contribute significantly to this process by intermittently
creating cool, dense layers of gas well above the photosphere where
dust grains can form and grow efficiently.

Pioneering work on the modelling of AGB star winds was done by Wood
(1979), \nocite{Wood79} focusing on the effects of shock waves, and
later by Bowen (1988), \nocite{Bowen88} introducing a parameterized
opacity to describe the dynamical effects of dust formation in the
circumstellar envelope. These early wind models where followed by
studies of carbon stars including time-dependent {(non-equilibrium)} grain growth
(e.g. Fleischer et al 1992; H\"ofner \& Dorfi 1997; Winters et al.
2000) which, despite being based on grey radiative transfer, allowed
to describe basic properties of heavily dust-enshrouded carbon
stars. In order to obtain reasonably realistic results for objects
with less optically thick envelopes, however, it is necessary to
combine frequency-dependent radiative transfer (including gas and
dust opacities) with time-dependent hydrodynamics and
non-equilibrium dust formation (cf. H\"ofner et al.~2003).

At a point where models of carbon stars are becoming quantitatively
comparable to observations as diverse as high-resolution IR spectra
(e.g. Nowotny et al. 2010) and spectro-interferometric measurements
(e.g. Sacuto et al. 2011), it is necessary to scrutinize a number of
underlying physical assumptions and approximations. In particular,
this concerns the detailed treatment of dust opacities that are at
the core of the wind mechanism and also have a direct influence on
observable properties. A common feature of most detailed dust-driven
wind models in the literature (including the first paper in this
series by Mattsson et al. 2010) is that dust opacities are computed
under the assumption that the grain sizes are small compared to the
relevant wavelengths (defined by the stellar flux distribution),
using the small-particle limit (SPL) of the Mie theory. In this limit, dust
opacities are fully determined by the amount of condensed material,
irrespective of grain sizes, which greatly simplifies the modelling
because an explicit knowledge of the actual grain size distribution in
each layer is not required. However, it has been shown that
grains may grow to sizes where the use of the SPL
is questionable (e.g., Gail \& Sedlmayr 1987, Winters et al. 1994,
1997, Mattsson et al.~2010).

The ongoing debate on the mass-loss mechanism of M-type AGB stars
has recently put the effects of grain size in these objects into
focus. Using detailed non-grey models, Woitke (2006) demonstrated
that silicate grains have to be virtually Fe-free in the wind
acceleration zone, which leads to insufficient radiative pressure caused by
absorption. Models by H{\"o}fner (2008) suggest scattering as a
possible solution: if conditions in the extended atmosphere allow
these grains to grow into the size range of about $0.1 - 1 \mu$m,
scattering becomes dominant over absorption by several orders of
magnitude, opening up the possibility of stellar winds driven by
scattering on virtually Fe-free silicate grains.

 \begin{figure}
 \includegraphics[width=9cm]{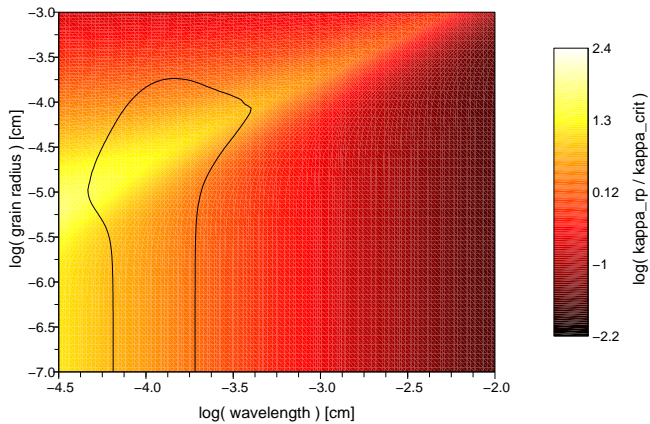}
 \caption{  \label{kappa_cont}
 Relevant dust opacity for radiation pressure
 {(combining effects of absorption and scattering,
 see Sect.~\ref{s_method})}
 as function of grain radius and wavelength,
 {computed from refractive index data for amorphous carbon}
 by Rouleau \& Martin (1991).
 The black contour shows the region where the flux-weighted
 monochromatic opacity exceeds the critical opacity,
 that is {\bf required in order for the radiation pressure to balance gravity}
 (see Eqs.~\ref{e_kappa_weight} and \ref{crit}, respectively),
 assuming a Planckian flux distribution with $T_{\rm eff} = 2700$~K
 and that 30\% of the carbon not bound in CO condense into carbon dust.
 }
 \end{figure}

In carbon stars, on the other hand, the effects of grain size are
expected to be less dramatic owing to the high absorption cross
sections of amorphous carbon grains. As shown in
Fig.~\ref{kappa_cont}, a wide range of particle sizes can contribute
to driving winds, which is different from Fe-free silicates, where small
particles are too transparent (cf. Fig.~1 in H{\"o}fner 2008). For
carbon grains with radii of about $0.1 - 1 \mu$m, however, the SPL
of the Mie theory may severely underestimate the
radiative pressure, with possible consequences for the wind properties.
The extensive model grid presented by Mattsson et al. (2010, hereafter 
Paper I) {shows that carbon grains in this size
range may be quite common,} in particular for conditions that allow
for efficient grain growth (high C abundance, low effective
temperatures, slow winds), {indicating} a potential
inconsistency with the underlying assumption of SPL opacities.

The main objective of this paper is to establish when the small
particle approximation can be applied, and to quantify the possible
errors that may occur in mass-loss rates, wind velocities, or
dust-to-gas ratios in cases where the dust particles grow beyond
this regime. For that purpose, we have implemented a generalized
description of dust opacities in our models, using actual mean grain
sizes and corresponding radiative cross sections that are valid
beyond the SPL (see Sect.~\ref{s_method}). From the
model grid in Paper~I (based on the SPL) we select
a subgroup of models that show large grains and/or slow winds, and
which we therefore can expect to be noticeably affected by the
assumptions about dust opacities. Re-computing these models with the
newly implemented, generalized treatment of grain opacities gives
estimates of the errors introduced by using the small particle
limit. It should be noted, however, that the extreme cases discussed
here are not necessarily representative of the majority of the wind
models in Paper~I, but that they rather highlight grain size as a
potentially critical property, and presumably give an upper limit of
the errors.

\section{Modelling method}\label{s_method}

The results discussed in this paper are based on dynamic atmosphere
and wind models that combine non-equilibrium dust formation and
frequency-dependent radiative transfer (taking both molecular and
dust opacities into account). The effects of stellar pulsation are
simulated by a variable inner boundary ('piston' with accompanying
luminosity variation) below the stellar photosphere. The general
modelling method has been described in detail by H\"ofner et al.
(2003) and Mattsson et al. (2007, 2010).

The new models presented here include a description of dust
opacities that is applicable to grains of arbitrary size, in
contrast to our earlier carbon star models, which used a simple limit-
case of the Mie theory -- valid for particles much smaller than the
relevant wavelengths only -- irrespective of the actual emerging
particle sizes. The more general method of computing radiative cross
sections can affect both the radiative energy transfer (temperature
structure) and the radiative pressure on dust grains, and,
consequently, the acceleration of the wind, if dust particles grow
beyond sizes where the SPL holds. Below,
we discuss the newly implemented description of dust opacities in
detail.

\subsection{Dust opacities: dependence on grain size}

A crucial quantity in the following discussion is the radiation
pressure efficiency factor $Q_{\rm rp}$, defined as the ratio of the
corresponding radiative cross-section, $C_{\rm rp}$, to the
geometric cross-section of a grain, i.e.,
\begin{equation}
  Q_{\rm rp} (a_{\rm gr},\lambda)
  \equiv {C_{\rm rp} (a_{\rm gr},\lambda) \over \pi\,a_{\rm gr}^2} \, ,
\end{equation}
assuming spherical grains with radii $a_{\rm gr}$. The cross
section determining radiative pressure 
is a combination of absorption and scattering cross-sections
($C_{\rm abs}$ and $C_{\rm sca}$, respectively), 
\begin{equation}
  C_{\rm rp} 
  = C_{\rm abs} + (1-g_{\rm sca})\,C_{\rm sca},
\end{equation}
with $g_{\rm sca}$ denoting the mean 
cosine of the scattering angle, where $g_{\rm sca}=1$ corresponds to
pure forward scattering \citep[see, e.g.,][]{Kruegel03}. These
quantities can be derived from refractive index data of relevant
grain materials using the Mie theory. With the definitions given above,
the opacity that determines the radiative pressure on an ensemble of
dust grains embedded in a gas -- with $\rho$ denoting the mass
density of the gas-grain mixture -- can be expressed as
\begin{equation}
\label{qint} \kappa_\lambda^{\rm dust} = \frac{\pi}{\rho}
\int_0^\infty a_{\rm gr}^2 Q_{\rm rp}(a_{\rm gr},\lambda)\,n(a_{\rm
gr})\, da_{\rm gr},
\end{equation}
where $n(a_{\rm gr})\, da_{\rm gr}$ is the number density of grains
in the size interval $da_{\rm gr}$ around $a_{\rm gr}$. By defining
  $Q_{\rm rp}' \equiv Q_{\rm rp} (a_{\rm gr},\lambda)/a_{\rm gr}$
and its grain-size average
\begin{equation}\label{tmq}
  \langle {Q'_{\rm rp}}\rangle
   = {\displaystyle \int_0^\infty Q'_{\rm rp}(a_{\rm gr},\lambda) \, a_{\rm gr}^3 n(a_{\rm gr})\, da_{\rm gr} \over
     \displaystyle \int_0^\infty a_{\rm gr}^3 n(a_{\rm gr})\, da_{\rm gr}},
\end{equation}
the opacity can be reformulated (without loss of generality) as
\begin{equation}\label{qout}
  \kappa_\lambda^{\rm dust} =
  \frac{\pi}{\rho} \, \langle Q'_{\rm rp} \rangle \int_0^\infty a_{\rm gr}^3 \,n(a_{\rm gr})\, da_{\rm gr},
\end{equation}
which is a more suitable form for the following discussion.

In our models the dust particles at distance $r$ from the stellar
center, at time $t$, are described in terms of moments $K_i (r,t)$
of the grain size distribution function $n(a_{\rm gr}, r, t)$, 
\begin{equation}
 K_i (r,t) \propto \int_0^\infty a_{\rm gr}^i\,n(a_{\rm gr},r,t)\,da_{\rm gr}
 \qquad (i = 0, 1, 2, 3) \, .
\end{equation}
It follows from this definition that $K_0$ is proportional to the
total number density of grains (the integral of the size
distribution function over all grain sizes), while $K_1$, $K_2$, and
$K_3$ are related to the average radius, geometric cross-section and
volume of the grains, respectively. The equations determining the
evolution of the moments $K_i (r, t)$ (including nucleation, grain
growth, and evaporation; cf. Gail \& Sedlmayr 1988, Gauger et
al.~1990) are described in detail in previous papers (see H{\"o}fner
et al. 2003 and references therein).

Regarding the computation of dust opacities, the integrals over
grain size in Eq.~(\ref{qout}) and in the denominator of
Eq.~(\ref{tmq}) are given by the moment $K_3$, while, in general,
the time-dependent local grain size distribution in each layer of
the model has to be known to evaluate the remaining
integral in Eq.~(\ref{tmq}), involving $Q'_{\rm rp}(a_{\rm
gr},\lambda)$.

 \begin{figure}[!ttt]
 \resizebox{\hsize}{!}{
 \includegraphics{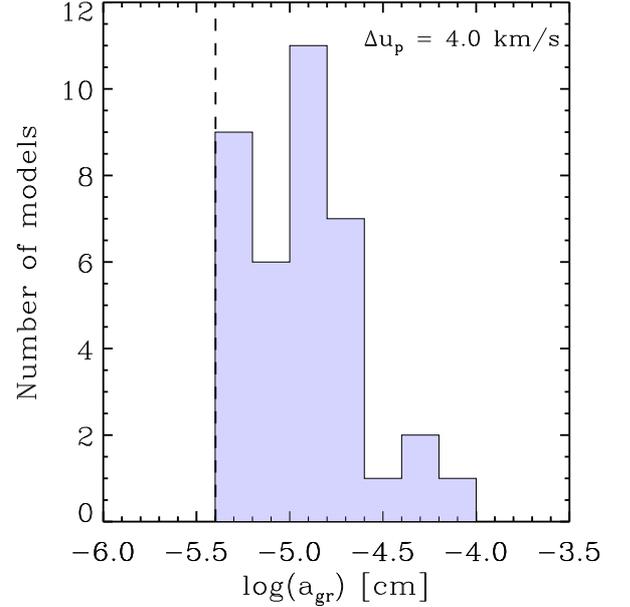}}
 \caption{  \label{agr_hist}
 Histogram of resulting mean grain sizes
 (derived from moment $K_1$ at the outer model boundary;
 see Eq.~(\ref{eq_a_mean}) and text)
 for wind-forming models
 taken from Paper~I, with
 $M_\star=1M_\odot$ and $\Delta u_{\rm p} = 4$ km s$^{-1}$,
 spanning a range of stellar luminosities, effective
 temperatures and carbon abundances (see Table~2 in Paper I).
 The vertical dashed line marks
 {a grain radius of $4 \cdot 10^{-6}\,$cm}
 where deviations in the opacity from the small-particle limit (SPL)
 may exceed 10$\%$ at wavelength $\lambda = 1\,\mu$m (see
 Fig.~\ref{smallpart}).
 }
 \end{figure}

 \begin{figure}
 \resizebox{\hsize}{!}{
 \includegraphics{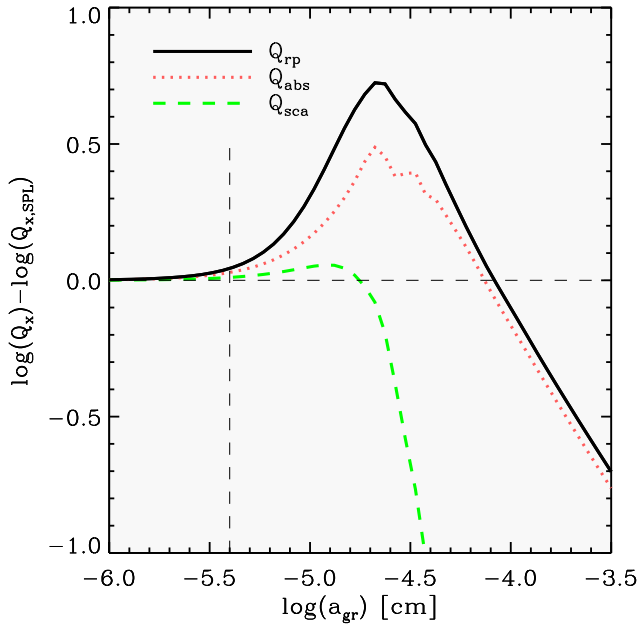}}
 \caption{  \label{smallpart}
 Radiative pressure efficiency factor $Q_{\rm rp}$ and its components
 $Q_{\rm abs}$ and $Q_{\rm sca}$,
 relative to the corresponding small-particle limit (SPL) values $Q_{\rm SPL}$,
 as functions of grain radius at $\lambda=1\,\mu$m (i.e., near the stellar flux maximum).
 Data are given for amorphous carbon dust, taken from Rouleau \& Martin (1991)
 and the $Q$'s are calculated using the Mie theory for spherical particles
 (programme BHMIE from Bohren \& Huffman 1983,
 modified by Draine, {\it www.astro.princeton.edu/draine/scattering.html}).
 The vertical dashed line marks
 {a grain radius of $4 \cdot 10^{-6}\,$cm}
 where deviations in the opacity from the SPL may exceed
 10$\%$ {at wavelength $\lambda = 1\,\mu$m}.
 }
 \end{figure}

\subsection{The small-particle approximation and its limitations}

In the limit case of particles that are much smaller than the relevant
photon wavelengths, i.e., $2 \pi a_{\rm gr} \ll \lambda$, however,
the problem of computing dust opacities becomes much simpler.
According to the Mie theory (see, e.g., Bohren \& Huffman, 1983), the
absorption and scattering efficiencies for small grains behave like
$Q_{\rm abs} \propto a_{\rm gr}$ and $Q_{\rm sca} \propto a_{\rm
gr}^4$. In this limit, absorption dominates over scattering,
implying that $Q_{\rm rp} \approx Q_{\rm abs}$, and, consequently,
that $Q_{\rm rp}' \approx Q_{\rm abs}/a_{\rm gr}$ becomes
independent of the grain size, making the integration in Eq.(\ref{tmq})
trivial. Therefore, provided that $2 \pi a_{\rm gr} \ll \lambda$
holds for all relevant wavelengths, the opacity can be reformulated
as
\begin{equation}
\label{spl} \kappa_\lambda^{\rm dust} =
  \frac{\pi}{\rho} \, Q'_{\rm abs}(\lambda) \int_0^\infty a_{\rm gr}^3\,n(a_{\rm gr})\, da_{\rm gr}
  \, \propto \, Q'_{\rm abs} \, \, K_3,
\end{equation}
This expression only depends on the total amount of material
condensed into dust (given by $K_3$). Consequently, explicit
knowledge of the grain size distribution is not required, which
greatly simplifies the modelling. Therefore, many models in the
literature, including those presented in Paper~I, have used the
SPL to describe dust opacities.

In view of the resulting mean grain sizes for the model grid
presented in Paper~I, however, it is necessary to investigate the
possible effects of size-dependent grain opacities beyond the small
particle limit on mass-loss properties of C-type AGB stars. A
comparison of Fig.\ref{agr_hist} (showing a histogram of typical
grain sizes in Paper~I) with Fig.\ref{smallpart} (showing the
deviations from small particle cross sections as a function of grain
size) demonstrates that typical grain sizes are in a range where the
small-particle approximation may lead to a considerable
underestimation of the radiative pressure.

\subsection{Dust opacities beyond the small-particle limit}

In principle, the size distribution function $n(a_{\rm gr},r,t)$ can
be reconstructed from the moments $K_i (r,t)$, allowing for a
general treatment of dust opacities (Eqs.~(\ref{qout}) and
(\ref{tmq})), but this involves a considerable computational effort,
well beyond the scope of this paper. Instead, we test the influence
of size-dependent dust opacities on AGB winds with several
descriptions that
\begin{itemize}
  \item do not require explicit knowledge of the size distribution,
  \item are based on the actual moments $K_i (r,t)$, and
  \item are not restricted to the small particle limit.
\end{itemize}
All these descriptions are in the general form of Eq.~(\ref{qout}),
which means that they are dependent on the total amount of material
actually condensed into grains in each layer at a given instant
(integral over grain size, proportional to $K_3$),
\begin{equation}
\label{kd_model} \kappa_\lambda^{\rm dust} \propto
  \, \langle Q'_{\rm rp} (r,t) \rangle \, \, K_3 (r,t) \, ,
\end{equation}
but they use different approximations for $\langle {Q'_{\rm rp}
(r,t)}\rangle$. A common feature of the new models discussed here is
that the size-average of $Q'_{\rm rp}$ (as defined in Eq.~\ref{tmq})
is approximated by the value of $Q'_{\rm rp}$ for an average grain
size, $\tilde{a}_{\rm gr} (r, t)$, in each layer, i.e.,
\begin{equation}
  \langle {Q'_{\rm rp}} (r,t) \rangle =
  Q'_{\rm rp} (\tilde{a}_{\rm gr}(r,t),\lambda) \, .
\end{equation}
More precisely, we consider the following cases:
\begin{itemize}
\item[(A)] Models where a fixed grain radius
  $\tilde{a}_{\rm gr}$
  is used when computing
  $Q'_{\rm rp} (\tilde{a}_{\rm gr},\lambda)$
  while dust formation (determining $K_3$) is modelled as usual.
  We study three cases, where $\tilde{a}_{\rm gr}$ is taken to be
  either deep in the SPL ($10^{-7}$ cm),
  or `optimised' such as to obtain maximum $Q_{\rm rp}$
  around the stellar flux maximum
    ($3.55\cdot 10^{-5}$ cm),
  or very large
    ($10^{-3}$ cm).
  In the `optimised' case, $Q_{\rm rp}$ is about four times higher
  compared to the SPL at $1\mu$m
  (see Fig.~\ref{smallpart}).
  Models with a fixed grain radius are considered mainly for reference.
\item[(B)] Models where $\tilde{a}_{\rm gr}$
  is a mean grain radius derived from
  the moments $K_i (r,t)$ (with $i = 1,2,3$),
  i.e.,
  \begin{equation}\label{eq_a_mean}
    \tilde{a}_{\rm gr} (r,t)
    \equiv \langle {a_{\rm gr}^i}\rangle^{1/i}
    = a_{\rm mon} \,\left({K_i\over K_0}\right)^{1/i},
  \end{equation}
  where $a_{\rm mon}$ is the monomer radius of the grain material.
  Which of the three cases gives the best approximation of the
  true $\langle {Q'_{\rm rp}}\rangle$ (defined in Eq. \ref{tmq}) is
  impossible to say without prior knowledge of the properties of
  $n(a_{\rm gr})$. Hence, we tested all three cases.
  The different mean radii $\tilde{a}_{\rm gr}$
  will be referred to as $K_1$, $K_2$ or $K_3$ mean, respectively.
  From a physical point of view, the $K_1$ mean is simply
  the mean grain radius resulting from the size distribution as such,
  while the $K_2$ and $K_3$ means represent grain radii corresponding
  to the mean grain surface and the mean grain volume, respectively.
\end{itemize}
In all cases the quantity $Q_{\rm rp} (\tilde{a}_{\rm
gr}(r,t),\lambda)$ was calculated using the Mie theory for spherical
particles of arbitrary size (using the programme BHMIE from Bohren \&
Huffman 1983, modified by Draine, {\it
www.astro.princeton.edu/draine/scattering.html}) and refractive
index data for amorphous carbon dust taken from Rouleau \& Martin
(1991).\footnote{Note that for all the descriptions of $\langle
{Q'_{\rm rp}}\rangle$ listed above the SPL of the Mie
theory is recovered if the assumed or actual grain sizes (in
type A or type B opacities, respectively) are much smaller than the
relevant wavelengths (defined by the stellar flux distribution),
i.e. $2 \pi a_{\rm gr} \ll \lambda$. In particular, opacities
of type A with $\tilde{a}_{\rm gr} = 10^{-7}\,$cm should be
directly comparable with Paper~I.}

For completeness sake, we also mention here that the
size dependence of the opacity relevant for determining the grain
temperature, i.e.~$Q_{\rm abs} (\tilde{a}_{\rm gr}(r,t),\lambda)$,
is treated in a similar way as the radiative pressure.

\section{Definitions and selection of models}

An average grain radius less than $4\cdot 10^{-6} $ cm is 
referred to throughout as "small" regarding radiative cross-sections.
This value corresponds to the lower limit of the size range
where the actual $Q_{\rm rp}$ (at wavelength $\lambda = 1\,\mu$m)
may deviate by more than 10\% from the value given by the small
particle limit (cf. Fig.~\ref{smallpart}). Below we will
also refer to dust opacities as being of type A or type B:, i.e.,
those with a fixed grain size (type A; for testing) and those
using varying characteristic grain sizes based on mean values
computed at each time step and spatial grid point throughout the
simulation (type B; see Sect. \ref{s_method} for details).

We present re-computations of two groups of
models selected from Paper I, using the modified version of our code
(see Sect. \ref{s_method}). The first group of models (numbers 1-12
in Tab.~\ref{data_spa}), referred to as "critical cases" in the
following, we expect to be significantly affected by including
grain-size effects. By "critical" we mean that the flux-mean dust
opacity is comparable to the critical opacity,
\begin{equation}
\label{crit}
\kappa_{\rm crit} = {4\pi c\, GM_\star\over L_\star},
\end{equation}
which corresponds to a ratio of unity for (outwards-directed)
radiative and (inwards-directed) gravitational acceleration for
stellar luminosity $L_\star$ and stellar mass $M_\star$ ($c$ and $G$
denote the speed of light and the constant of gravity,
respectively). In other words, critical cases are defined by a
situation where radiation pressure on dust is close to the value
required for balancing gravity. In practice this corresponds to
models with slow winds in which the dust grains have time to grow
larger than usual, and models near some mass-loss threshold in
stellar parameter space for which a slight increase/decrease of
$Q_{\rm rp}$ could enable or prevent wind formation.

In addition, a "control group" consisting of 12 models with strong,
well-developed outflows that also show relatively large average
grain sizes (models 13-24 in Table~\ref{data_spa}) was selected and
recomputed for comparison. These models are producing big dust grains
according to the definition above (see Fig. \ref{agr_hist}), but the
effects of grain-size dependent opacities may not be that
significant, because the momentum transfer efficiency (from radiation
to dust and gas) in these cases is near the theoretical maximum,
i.e., the single-scattering limit, which corresponds to the
mass-loss rate, $\dot{M} \sim {L_\star\,u_{\rm out} / c}$, where
$L_\star$ is the luminosity, $u_{\rm out}$ is the flow speed over
the outer boundary and $c$ is the speed of light. The optical depth
of the wind is high
and the wind speed cannot be much affected by an increase of the
radiative pressure efficiency factor $Q_{\rm rp}$.

\section{Results and discussion}

\subsection{Basic tests and constraints}

To test the modified code, we have tried to replicate the results
from Paper I by adopting a small fixed grain radius $a_{\rm gr} =
10^{-7}$ cm, when calculating $Q'_{\rm rp}$, which should be well
within the small particle regime. The average mass-loss rates,
wind speeds and mean degrees of dust condensation that we obtained
are indeed almost exactly the same as in Paper I, which
indicates that the modified code is working properly.

In the opposite limit, i.e., when the particles are much larger than
the wavelengths under consideration, $Q_{\rm rp}$ approaches a
constant value (see, e.g., Fig.~11 in Paper I), and, consequently,
$Q'_{\rm rp} \propto 1/a_{\rm gr}$. From Eq.~(\ref{qout}) it can
therefore be deduced that for a fixed total amount of dust material
per volume (represented by the integral over grain size),
$    \kappa_{\lambda}^{\rm dust} \propto 1 / a_{\rm gr}$
for $2 \pi a_{\rm gr} \gg \lambda$. Because the total amount of grain
material per volume is limited by the availability (abundances) of
the constituting chemical elements, there is a limiting maximum
grain-size where the flux-mean opacity necessarily drops below the
critical opacity $\kappa_{\rm crit}$ and radiative pressure alone
cannot overcome gravity.

Fig.~\ref{kappa_cont} illustrates the dependence of the dust opacity
on both grain size and wavelength, and shows an estimate of which
grain sizes will be relevant for driving winds. The colour scale
represents the quantity $\kappa_{\lambda}^{\rm dust}/\kappa_{\rm
crit}$ and the black contour marks the region where the
monochromatic flux-weighted opacity
\begin{equation}\label{e_kappa_weight}
  \tilde{\kappa}_{\rm rp}(\lambda, a_{\rm gr}) =
  \kappa_{\lambda}^{\rm dust} \, \, {\pi \lambda B_\lambda(T_{\rm eff})\over \sigma T_{\rm eff}^4}
\end{equation}
exceeds the critical opacity $\kappa_{\rm crit}$, assuming a
Planckian flux distribution $B_\lambda$ with $T_{\rm eff} = 2700$~K
and that 30\% of the carbon not bound into CO is condensed into
grains (with a free carbon abundance $\varepsilon_{\rm
C}-\varepsilon_{\rm O} = 3.3\cdot 10^{-4}$, $M_{\ast} =
1\,M_{\odot}$, $L_{\ast} = 7000\,L_{\odot}$ and $\sigma$ denoting
the Stefan-Boltzmann constant). An upper limit in grain sizes
relevant for driving a wind is clearly apparent from this plot.
Adopting a large fixed grain radius ($a_{\rm gr} = 10^{-3}$ cm)
should therefore prevent the formation of dust-driven outflows, which
is confirmed by our detailed models.

\subsection{Recomputed models with optimized type A opacities}

By "optimizing" the grain radius used in the opacities for
maximum $Q_{\rm rp}$, we obtain much faster winds and higher
mass-loss rates for the critical cases (Models 1-12) compared to
using the SPL (see Tables \ref{data_spa} \&
\ref{data_opt} and Fig. \ref{agr_spa_Kopt}, left column). Four out
of five cases where SPL models have no
resultant wind do indeed have a considerable outflow when using an
optimized $Q_{\rm rp}$. which is roughly a factor of five higher
compared to the SPL value.

The actual average grain radius derived from the moment equations
for dust formation, i.e., $\langle a_{\rm gr}\rangle = a_{\rm
mon}\,K_1/K_0$, is smaller for models with maximized type A dust
opacities compared to SPL models (see Fig. \ref{agr_spa_Kopt}).
This is a consequence of dust grains having less time to grow; the
flow is generally faster and therefore the grains pass through the
dust formation zone in a shorter time. In general the mean degrees
of dust condensation $\langle f_c\rangle$ are lower, which makes the
"dust-loss rates" several times lower.

The "control group" of models (13-24) with strong, well-developed
dust-driven outflows (when using the SPL) are, as expected, not much
affected by maximizing the dust opacity, as far as mass-loss rate
and wind velocity are concerned. Those are also models that
presumably are quite representative of the majority of
wind-producing models in Paper I. 

\subsection{Recomputed models with type B opacities}

\begin{figure*}[!ttt]
 \resizebox{\hsize}{!}{
 \includegraphics{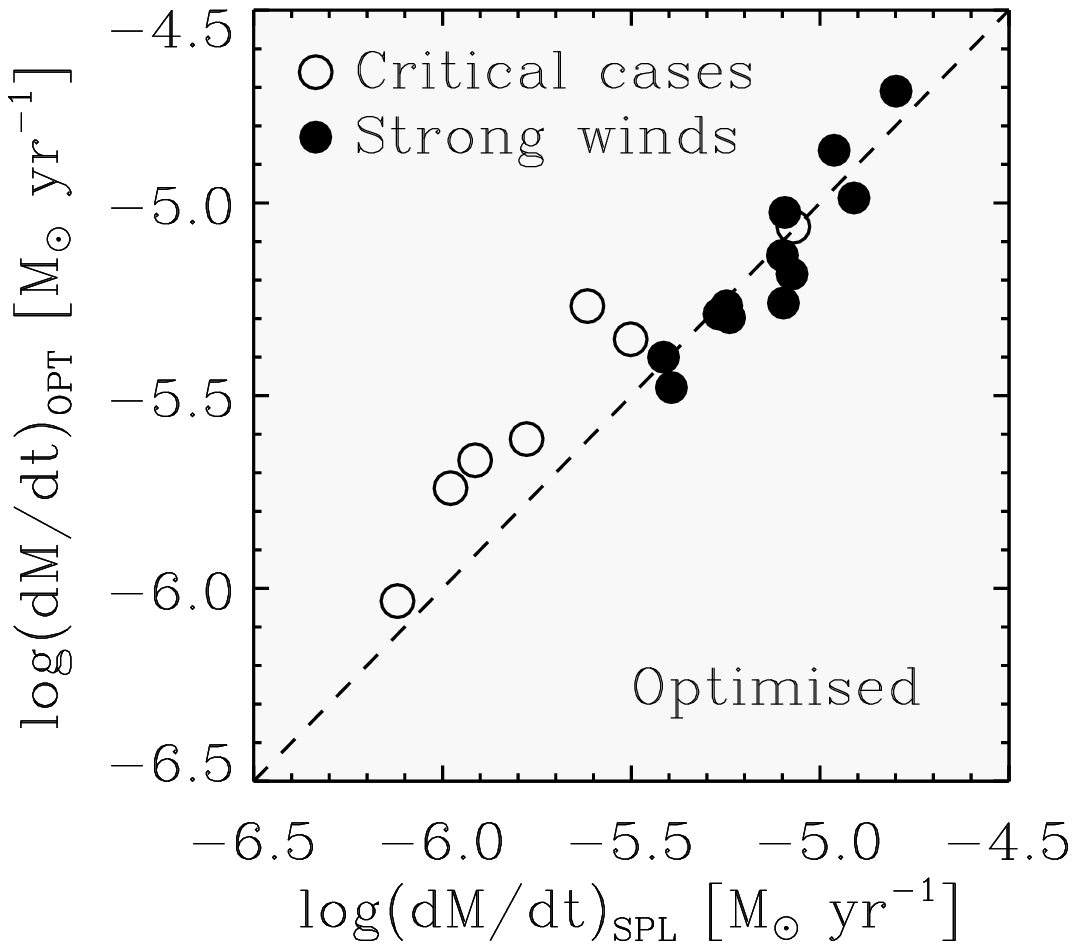}
 \includegraphics{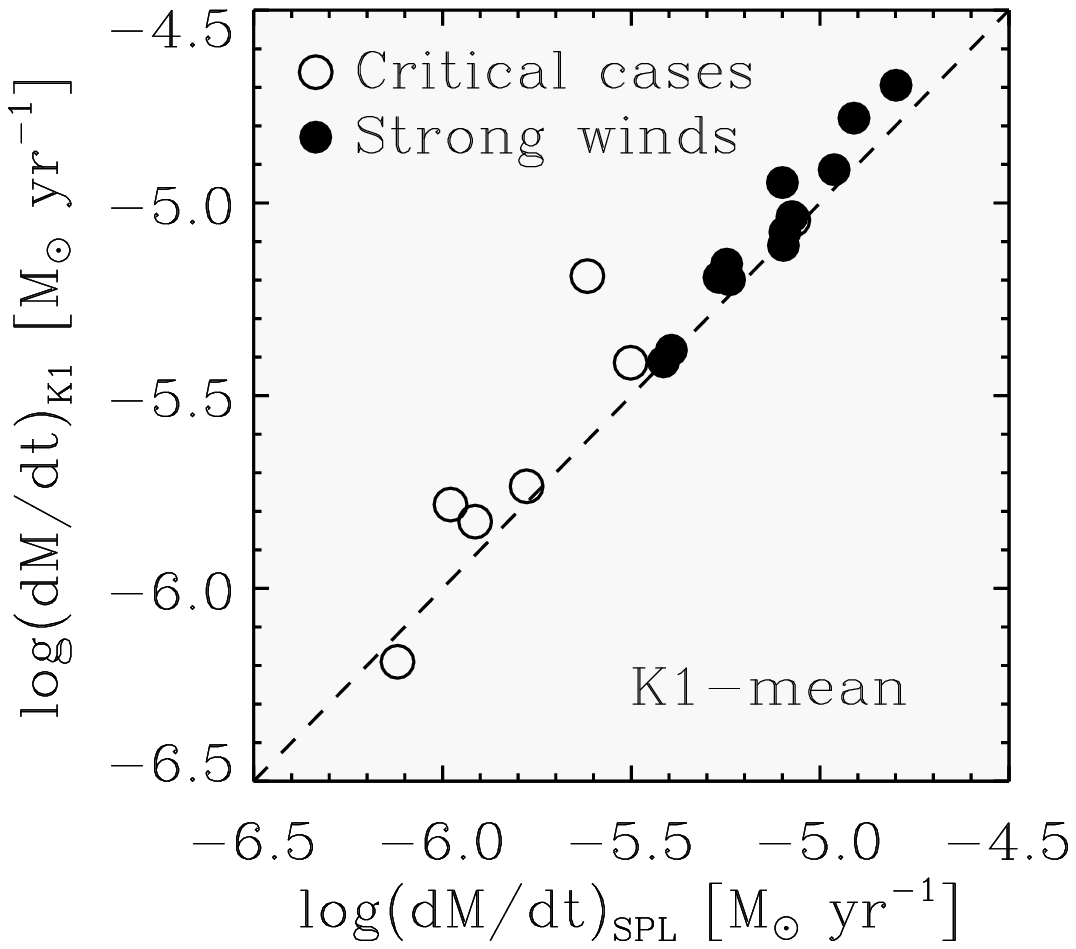}
 \includegraphics{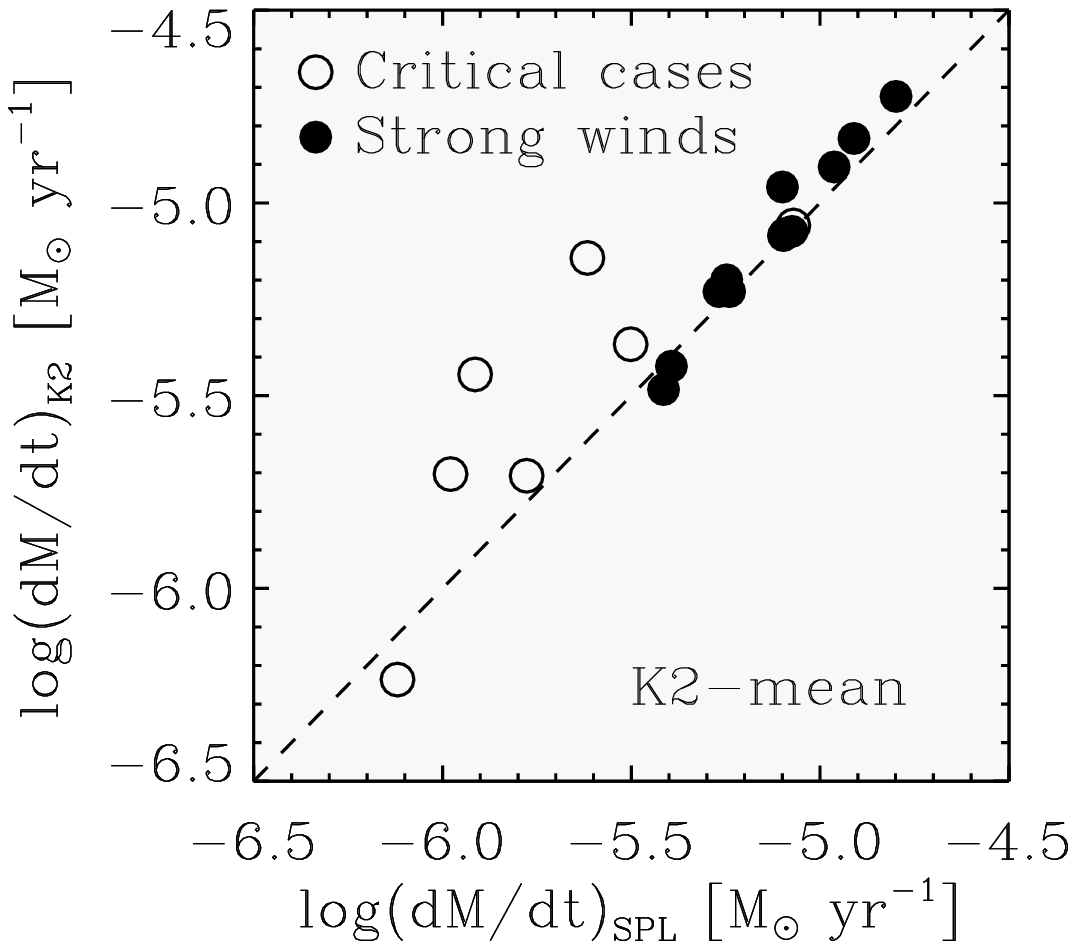}
 \includegraphics{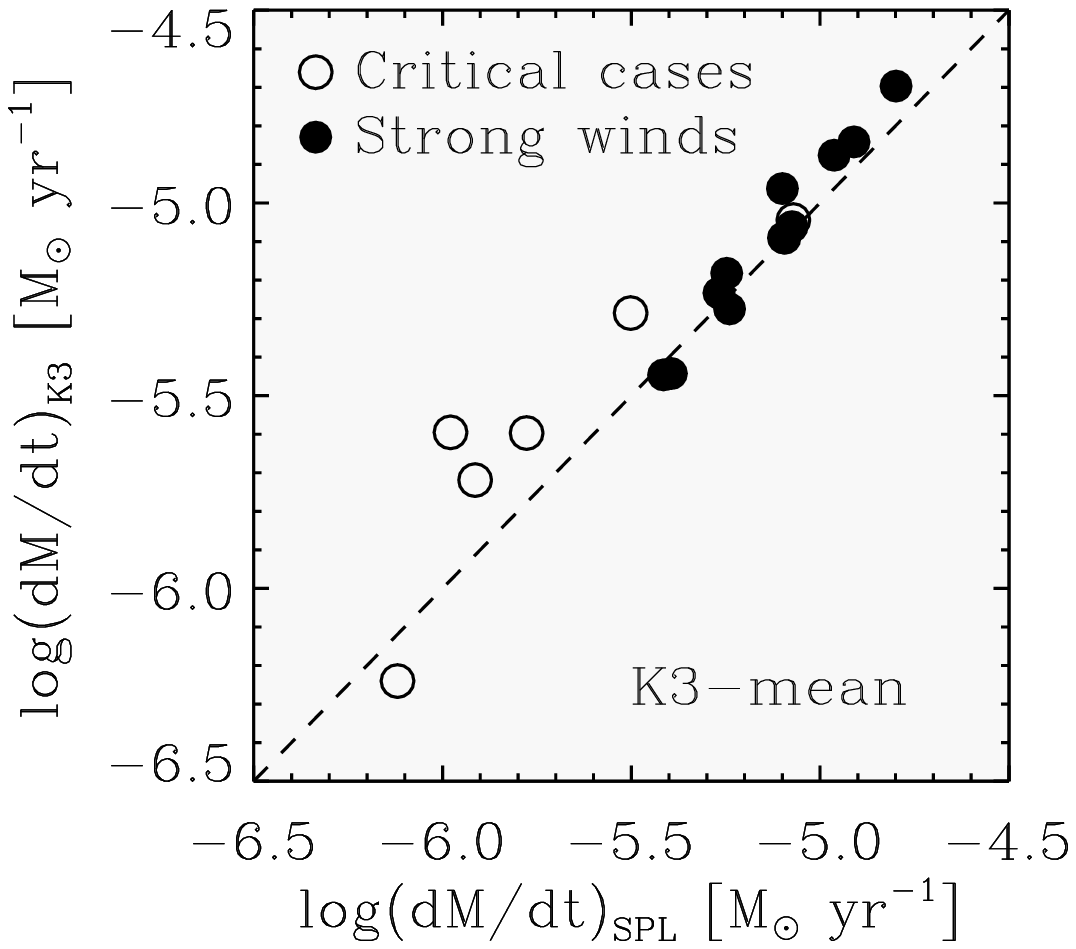}}

 \resizebox{\hsize}{!}{
 \includegraphics{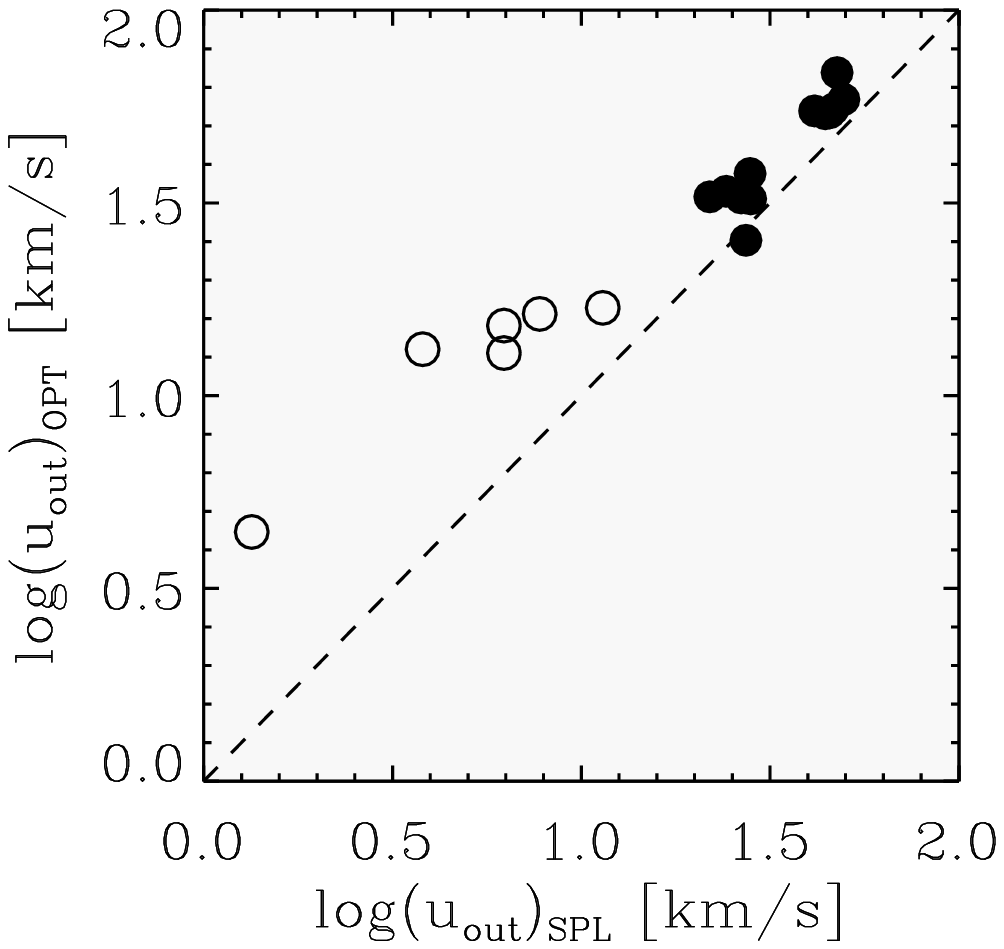}
 \includegraphics{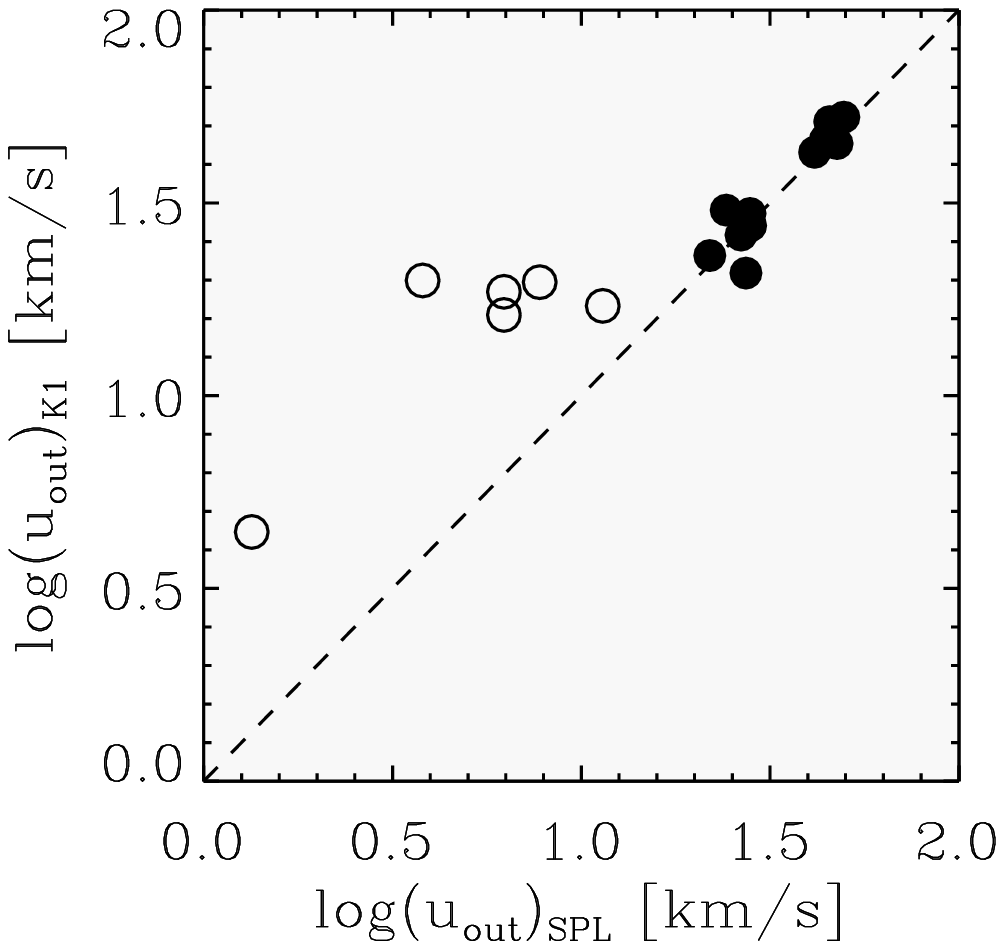}
 \includegraphics{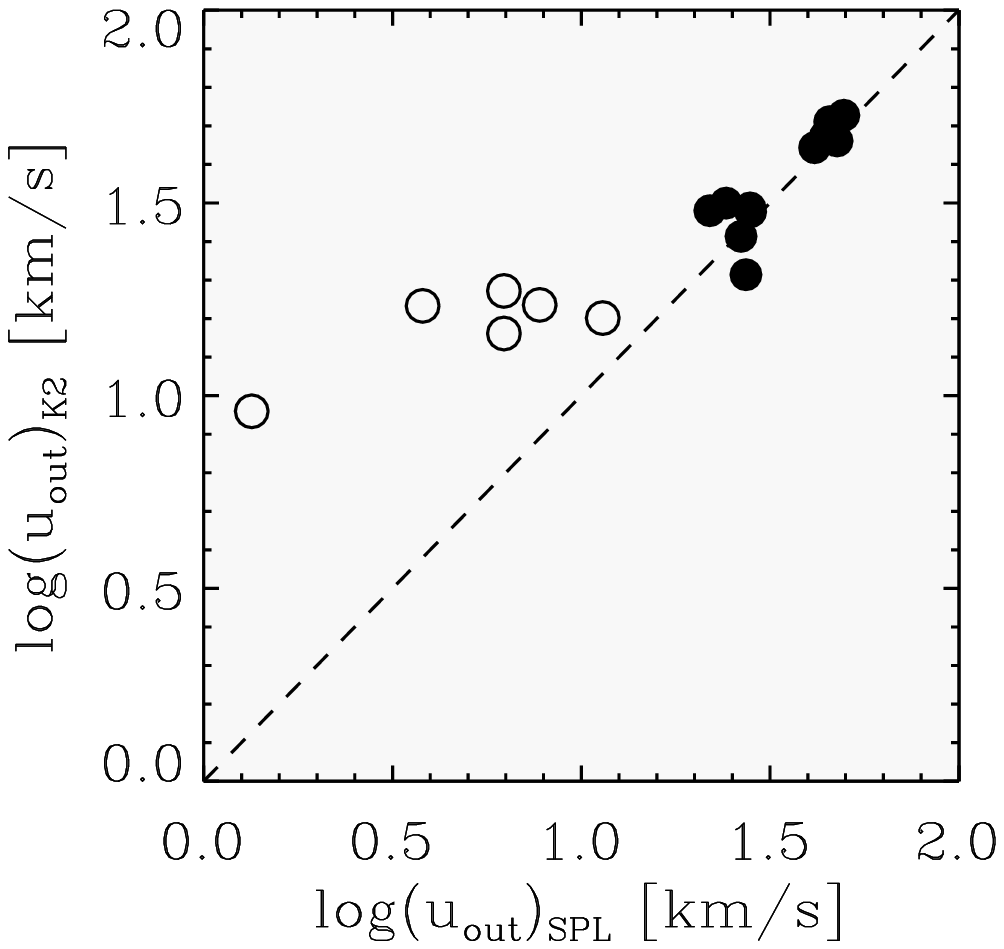}
 \includegraphics{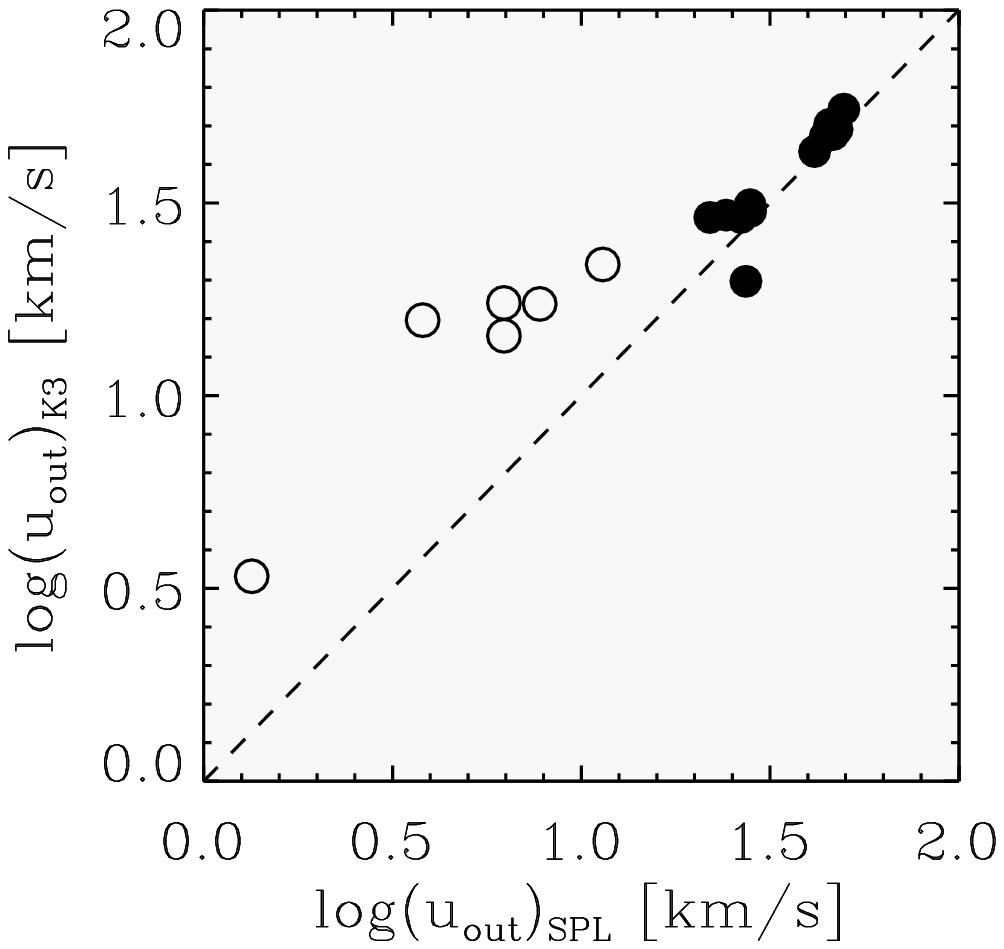}}

 \resizebox{\hsize}{!}{
 \includegraphics{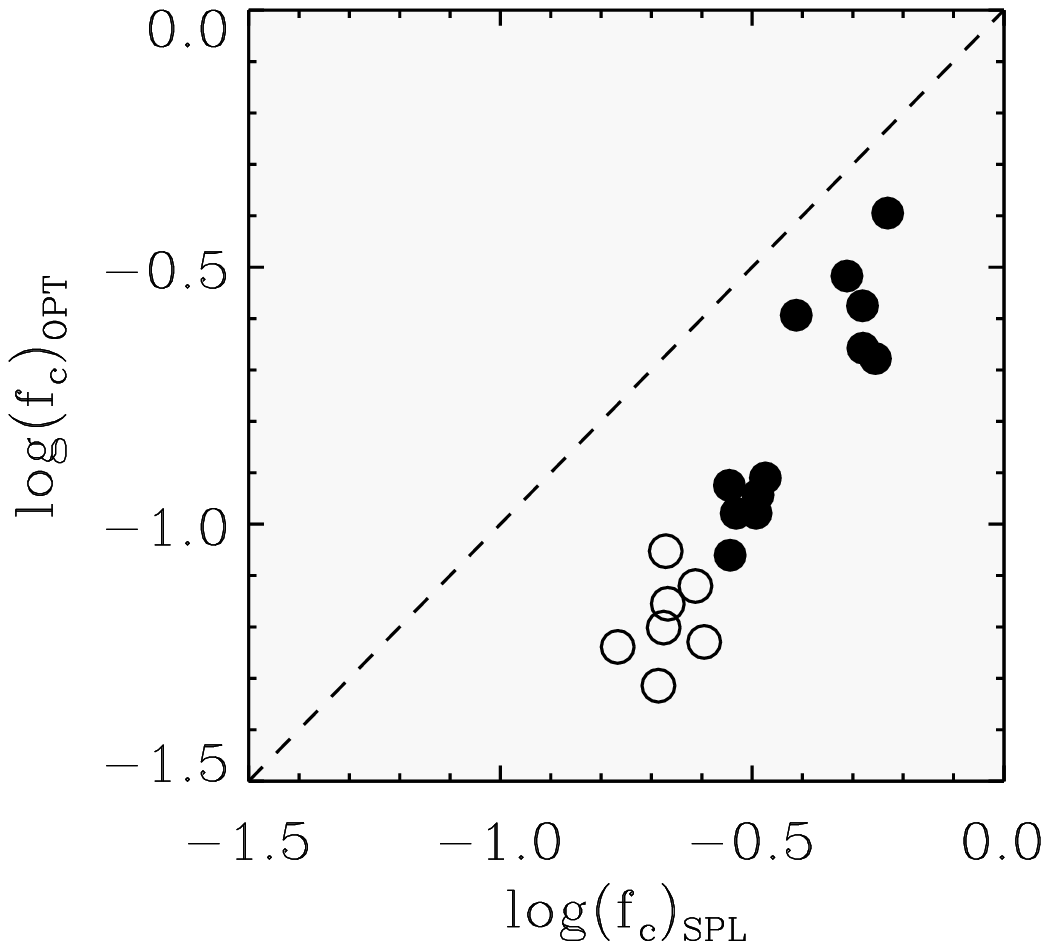}
 \includegraphics{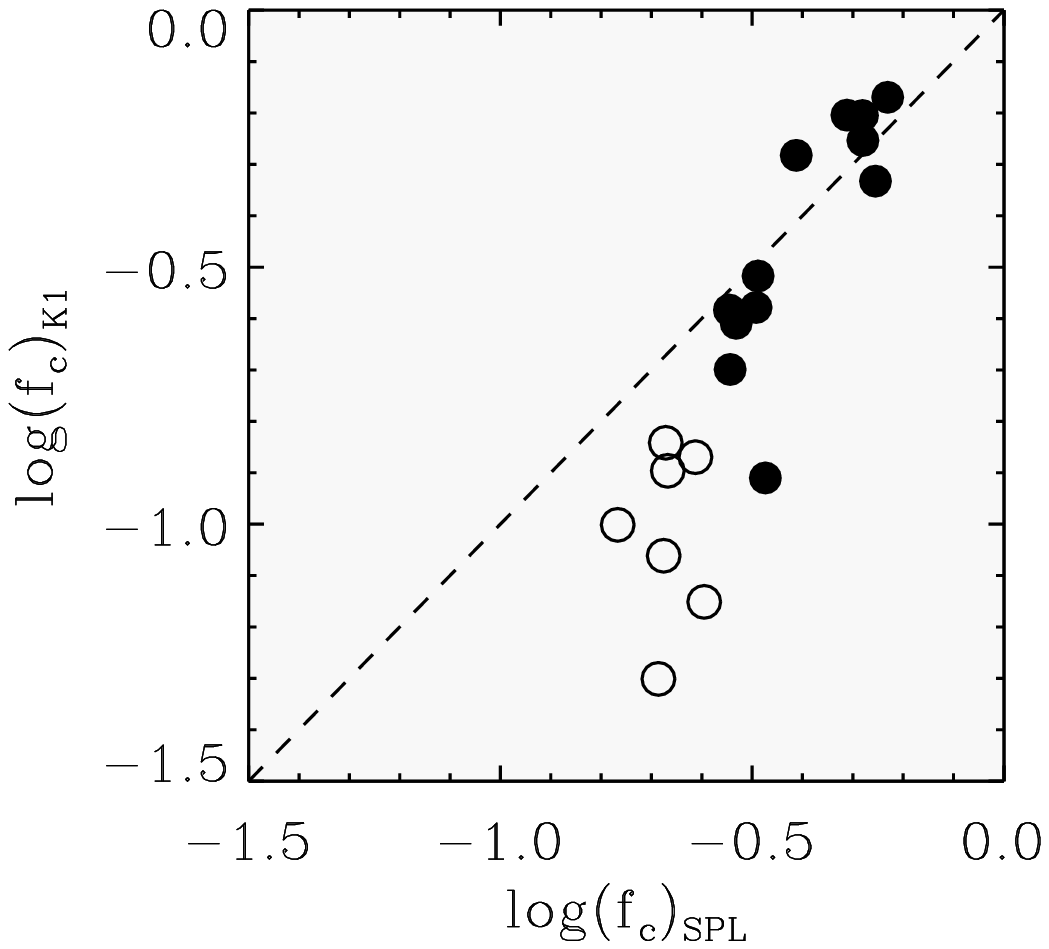}
 \includegraphics{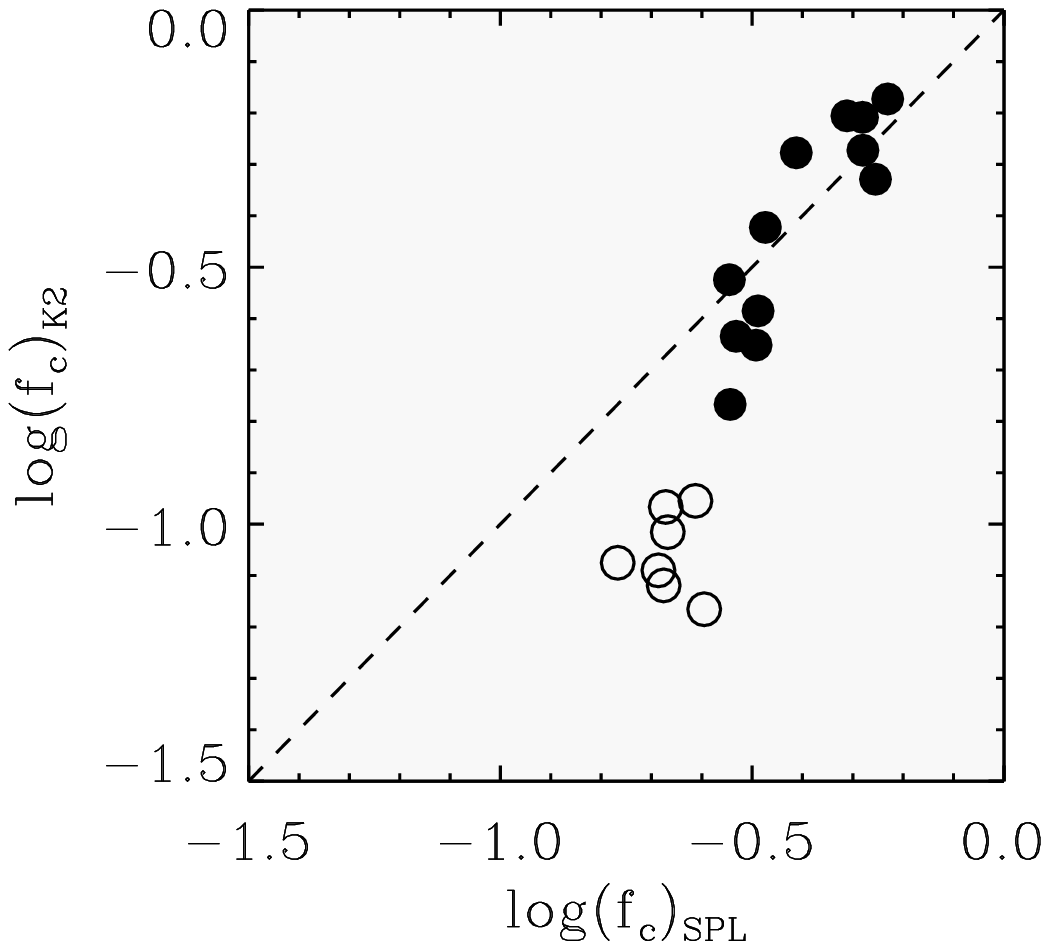}
 \includegraphics{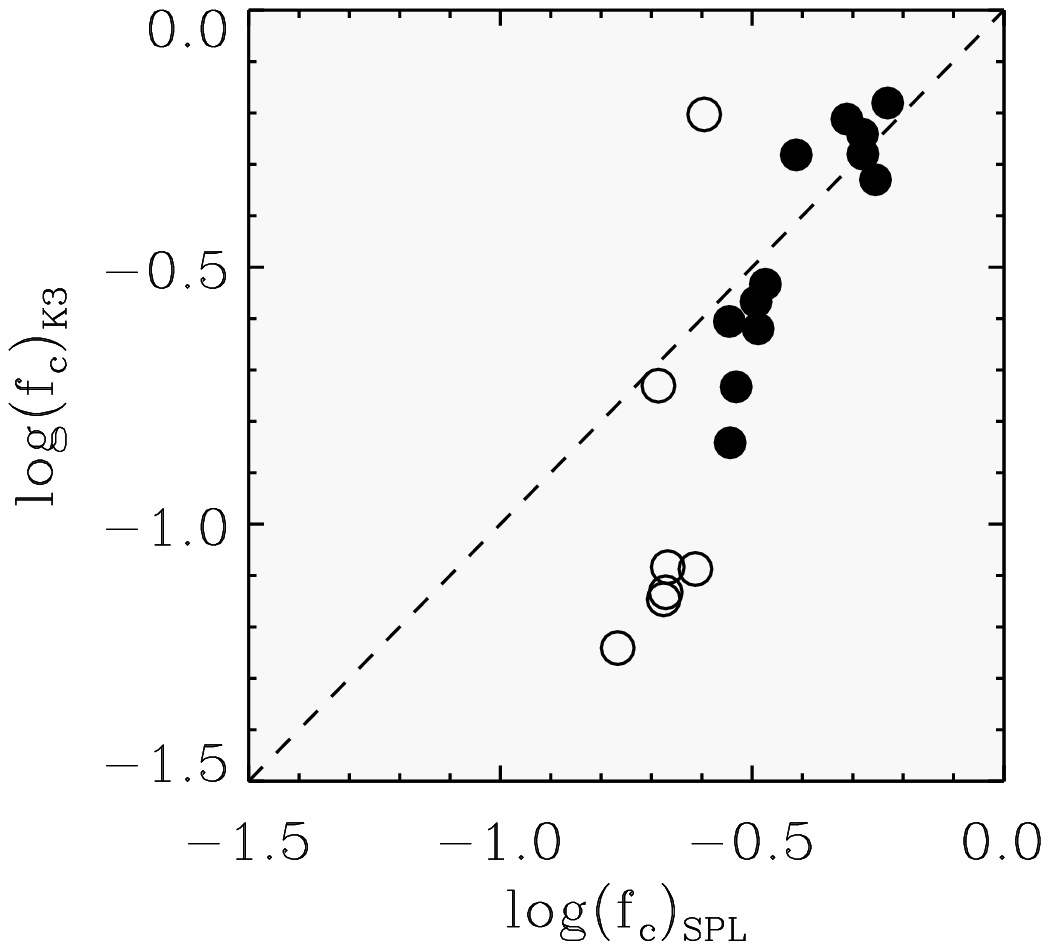}}

 \resizebox{\hsize}{!}{
 \includegraphics{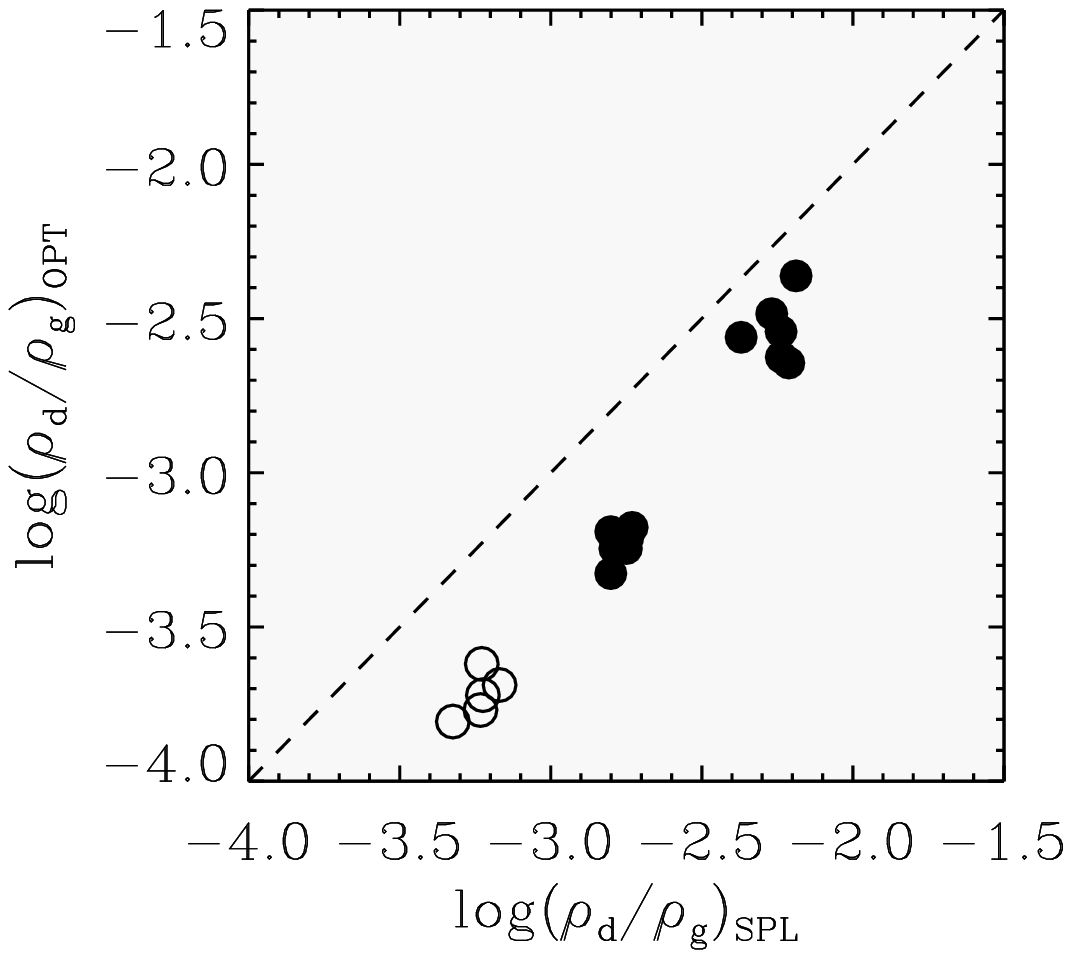}
 \includegraphics{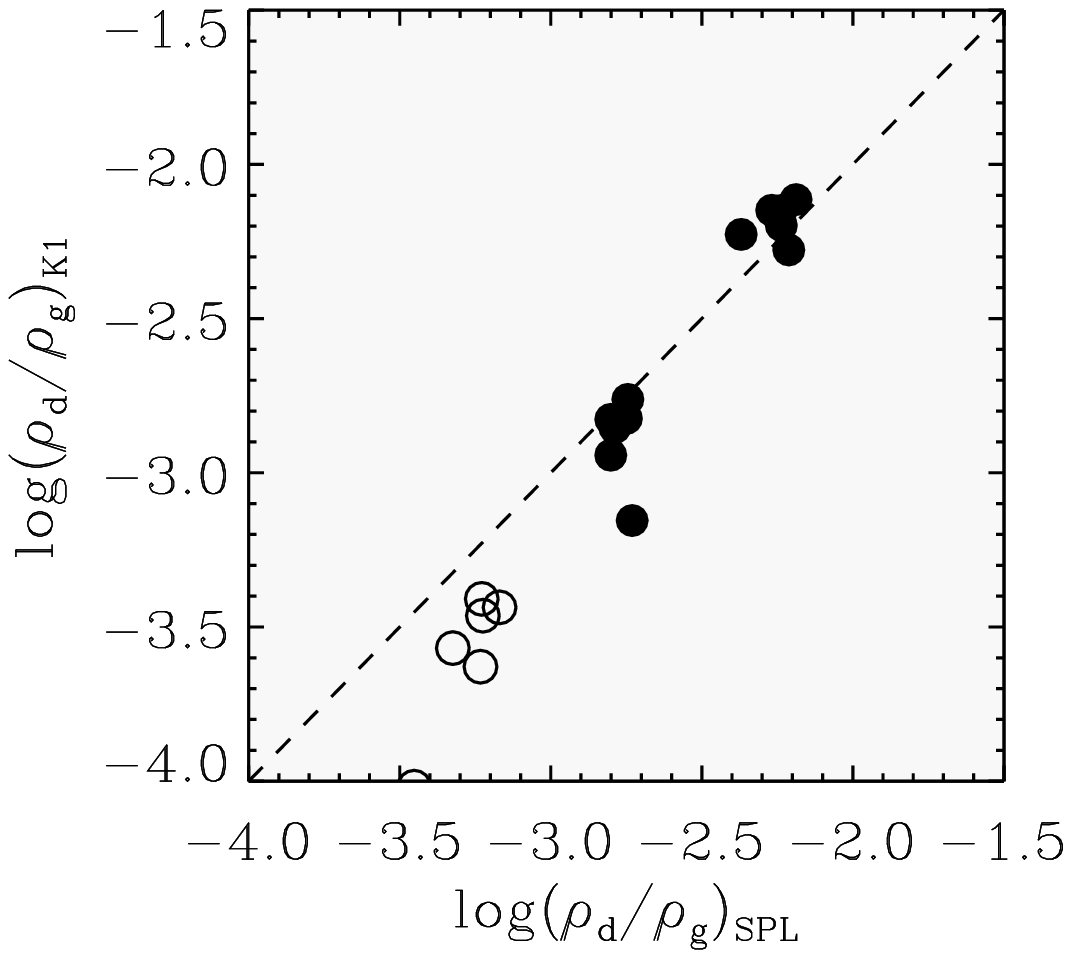}
 \includegraphics{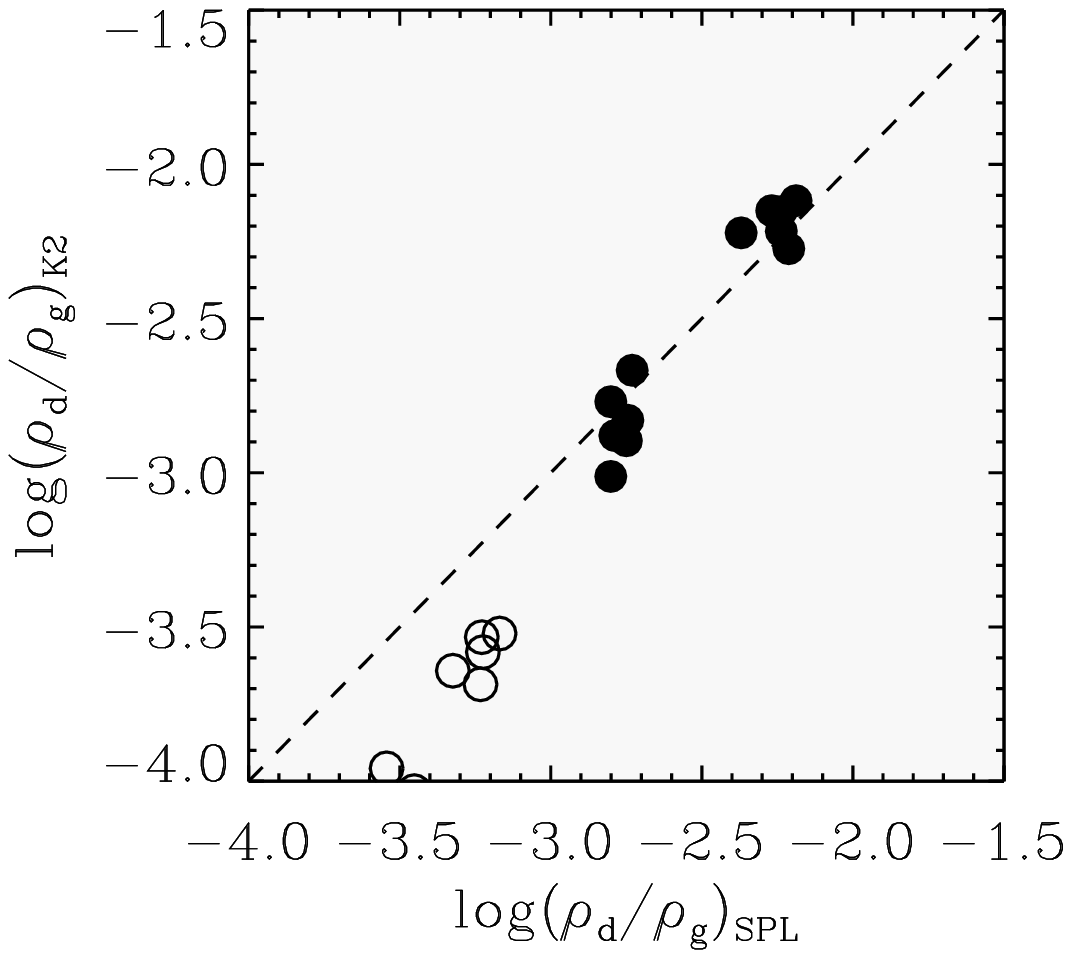}
 \includegraphics{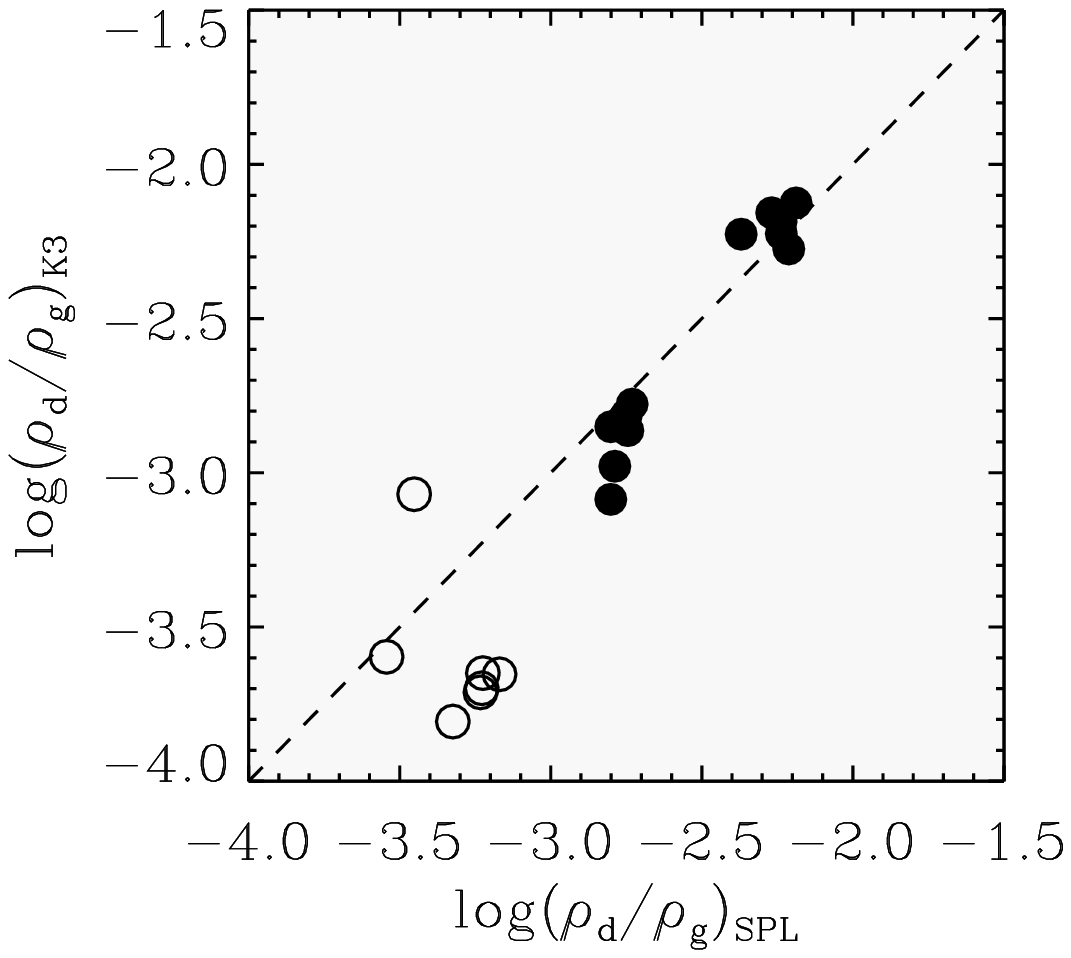}}

  \resizebox{\hsize}{!}{
 \includegraphics{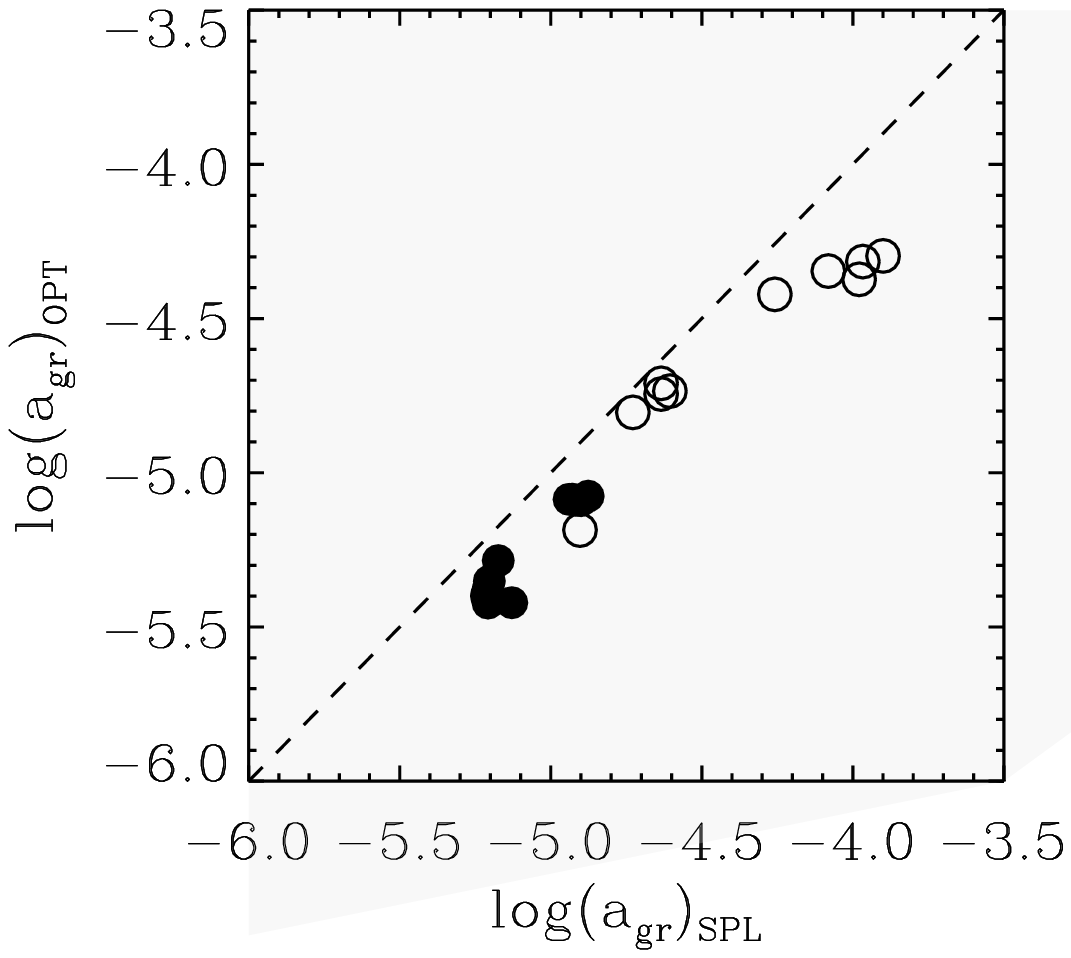}
 \includegraphics{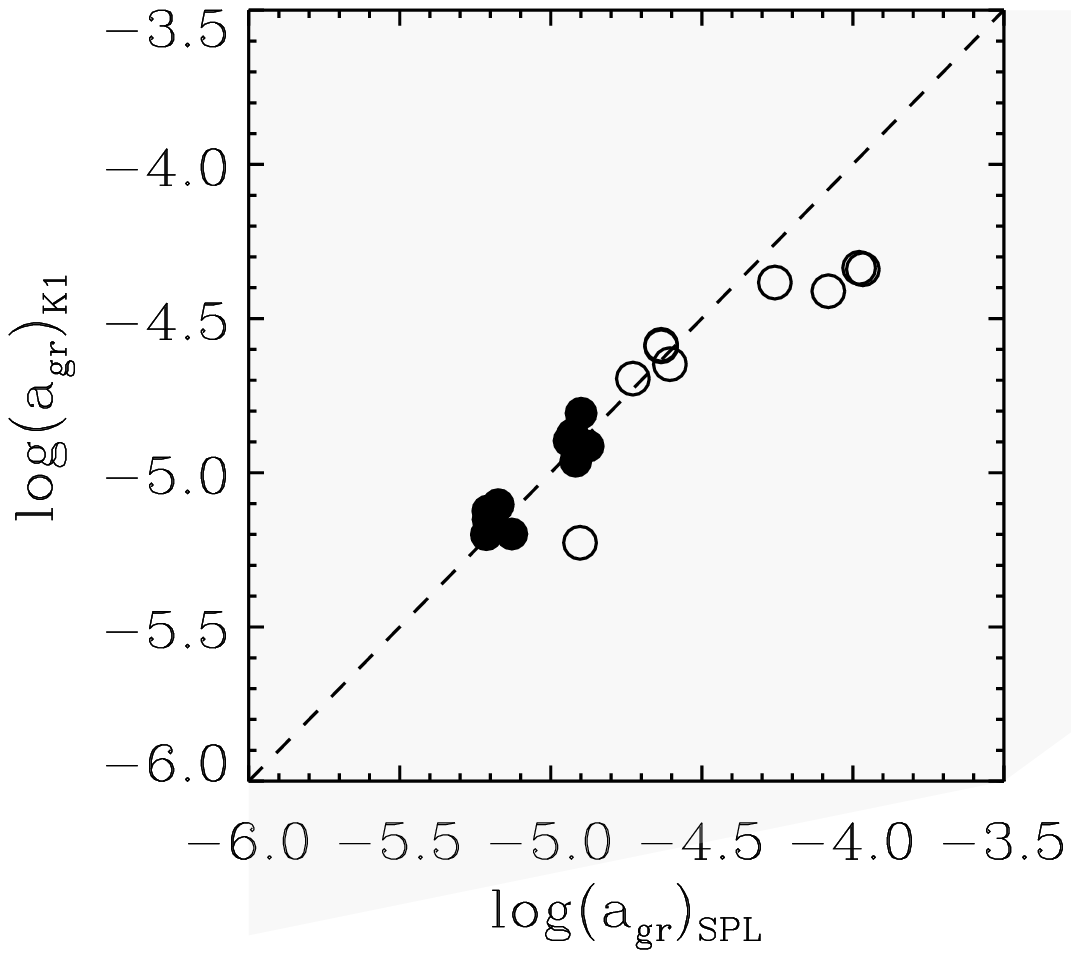}
 \includegraphics{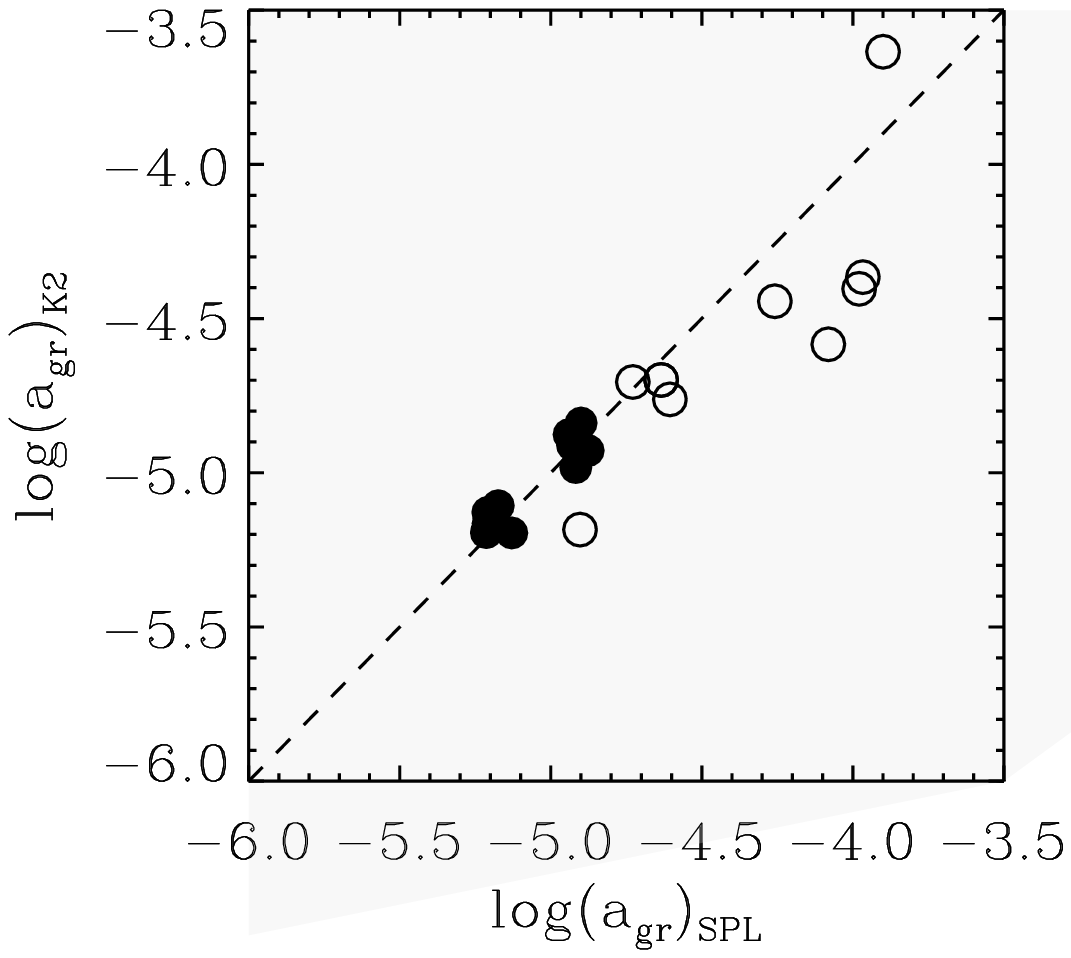}
 \includegraphics{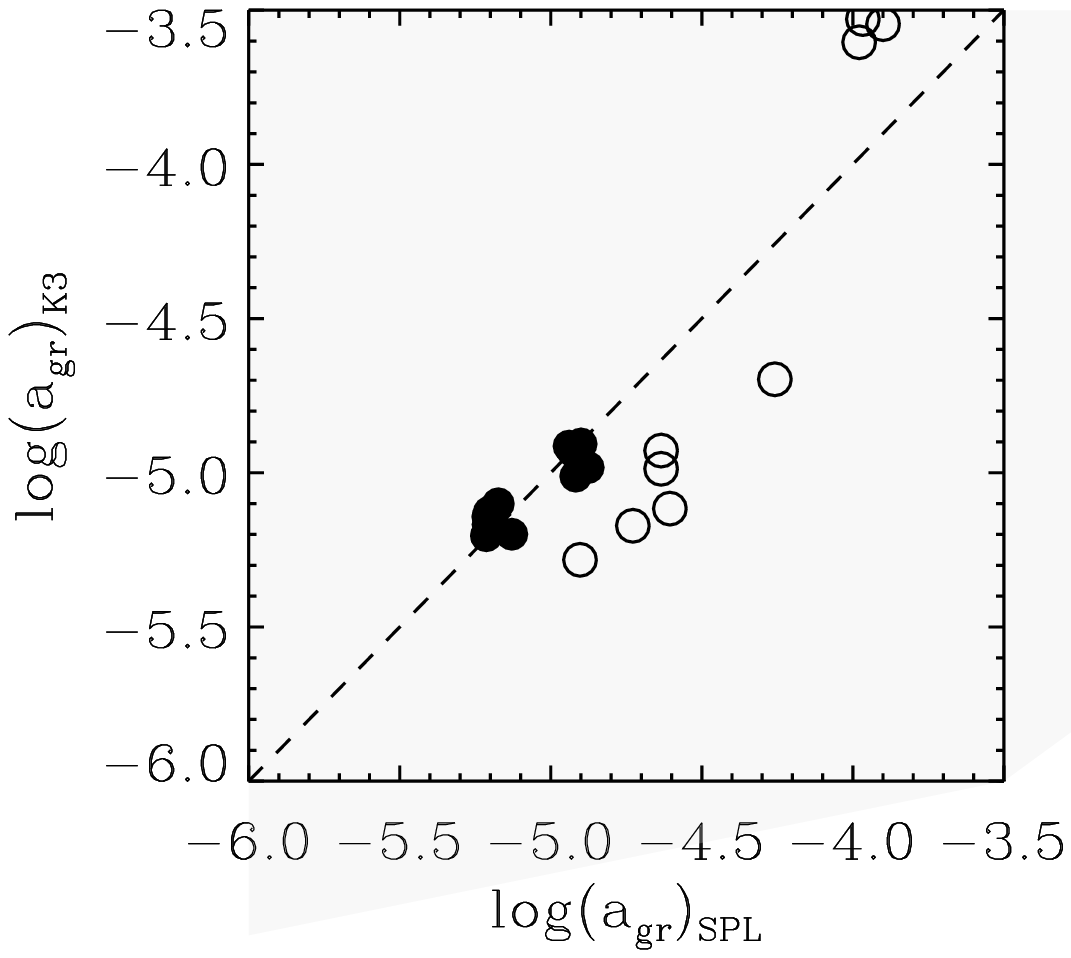}}

 \caption{From top to bottom:
 Mass-loss rates, wind speeds, mean degrees of dust condensation,
 dust-to-gas ratios and mean grain radii
 for models with type A opacities ("optimized $Q_{\rm rp}$", first column)
 and type B opacities
 (using actual grain sizes based on the moments
 $K_1$, $K_2$ and $K_3$; columns 2, 3 and 4, respectively)
 vs. the corresponding quantities in SPL models.
 The dashed lines show the case of equal values.
 \label{agr_spa_Kopt}}

 \end{figure*}

In reality, not all grains can obtain the same radii, and
especially not the radius that would maximize $Q_{\rm rp}$. The
models with type B dust opacities take into account that $Q_{\rm
rp}$ changes in time and space owing to variations of the grain-size
distribution, by using a value of $\langle Q'_{\rm rp} (r,t)
\rangle$ based on different moments of the grain-size distribution
(cf. Sect.~\ref{s_method}). For the critical-case models, this
results in much faster wind speeds and more intense mass loss
compared to using the SPL (see Fig. \ref{agr_spa_Kopt}). The
"control group" of models with strong winds are again much less
affected in these respects.

The results are generally quite similar to those of the models with
maximized type A opacities, which suggests that the average grain
size tends to be such that $Q_{\rm rp}$ is actually {\it close to
maximized} in the relevant part of the spectrum.
The average grain radii are in many cases larger than according to
type A models, but still smaller than the grain radii inferred from
the $K_1$ moment of the corresponding SPL models in Paper I.
The fact that the grain radii tend to be smaller when using the
generalized description of dust opacities does not mean that we are
approaching the small-particle region again (where the SPL holds
exactly), but reflects the effects of self-regulation in the
wind mechanism. When dust grains grow beyond the SPL regime,
the radiative acceleration becomes more efficient and they are then
likely to move away from the dust formation zone faster, which
means that they cannot continue to grow. If the momentum transfer
from the radiation field to the dust-gas mixture is not
sufficient to sustain an outflow, the dust grains may on the other
hand continue to grow, which means that small particles (experiencing
too little radiation pressure) may grow until they reach optimal size.

The mean degree of dust condensation $\langle f_c\rangle$ is
typically much lower for critical-case models with type B
opacities than for the corresponding SPL models of Paper I, but
slightly larger than for models with type A opacities (see Fig.
\ref{agr_spa_Kopt}). There is one critical $K_3$ mean model (Model
5) that stands out from the rest and shows a $\langle f_c\rangle$
value that is several times higher than for the corresponding models
with maximized type~A opacities, or when $\langle Q'_{\rm
rp}\rangle$ is computed with the $K_1$ and $K_2$ mean grain
radii (again, see Fig. \ref{agr_spa_Kopt}). The reason is that
this particular model shows no net outflow, which means that grain
growth is not stopped by falling densities as would be the case in a
wind.

 \begin{figure*}
 \resizebox{\hsize}{!}{
 \includegraphics{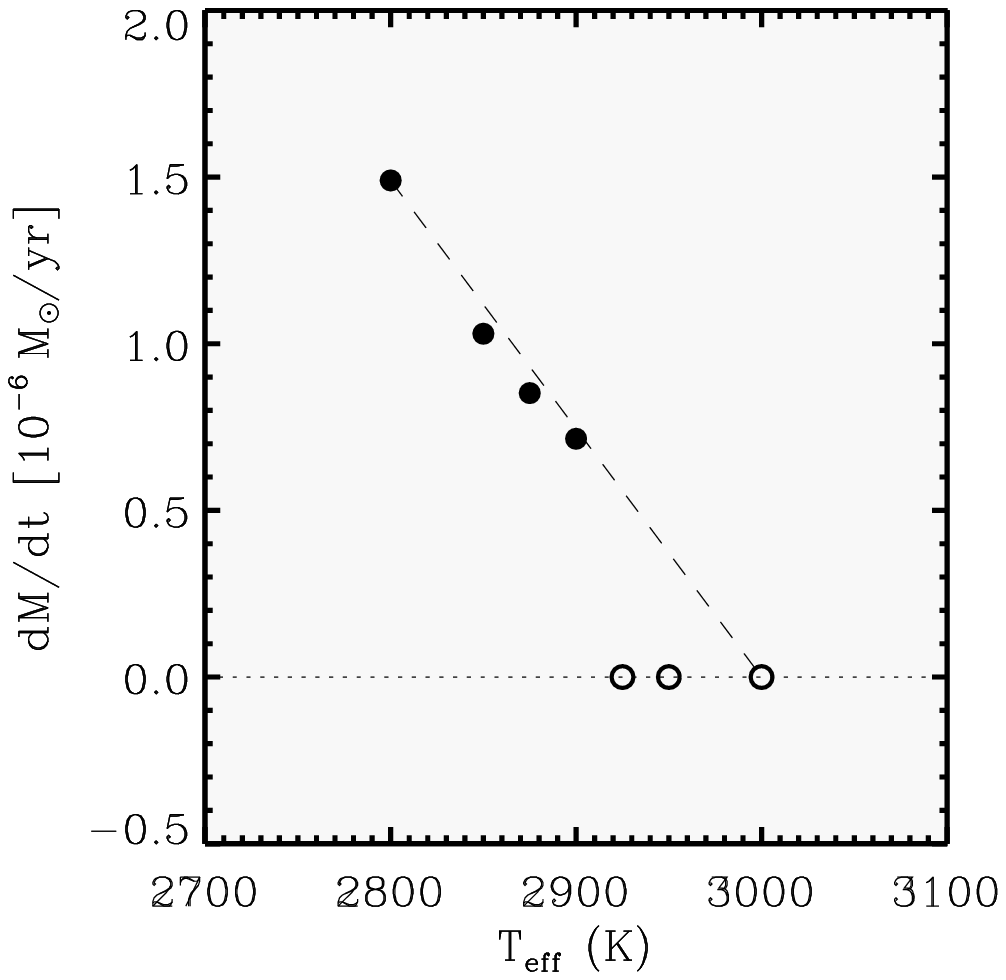}
 \includegraphics{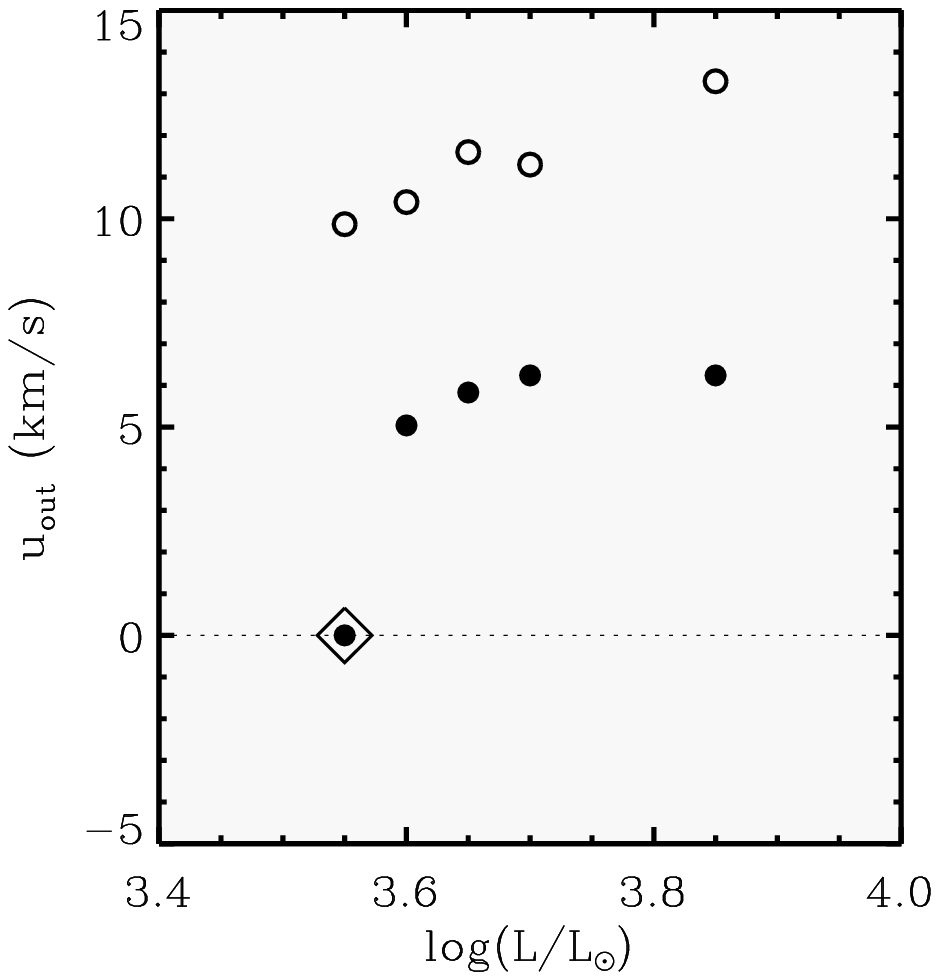}
 \includegraphics{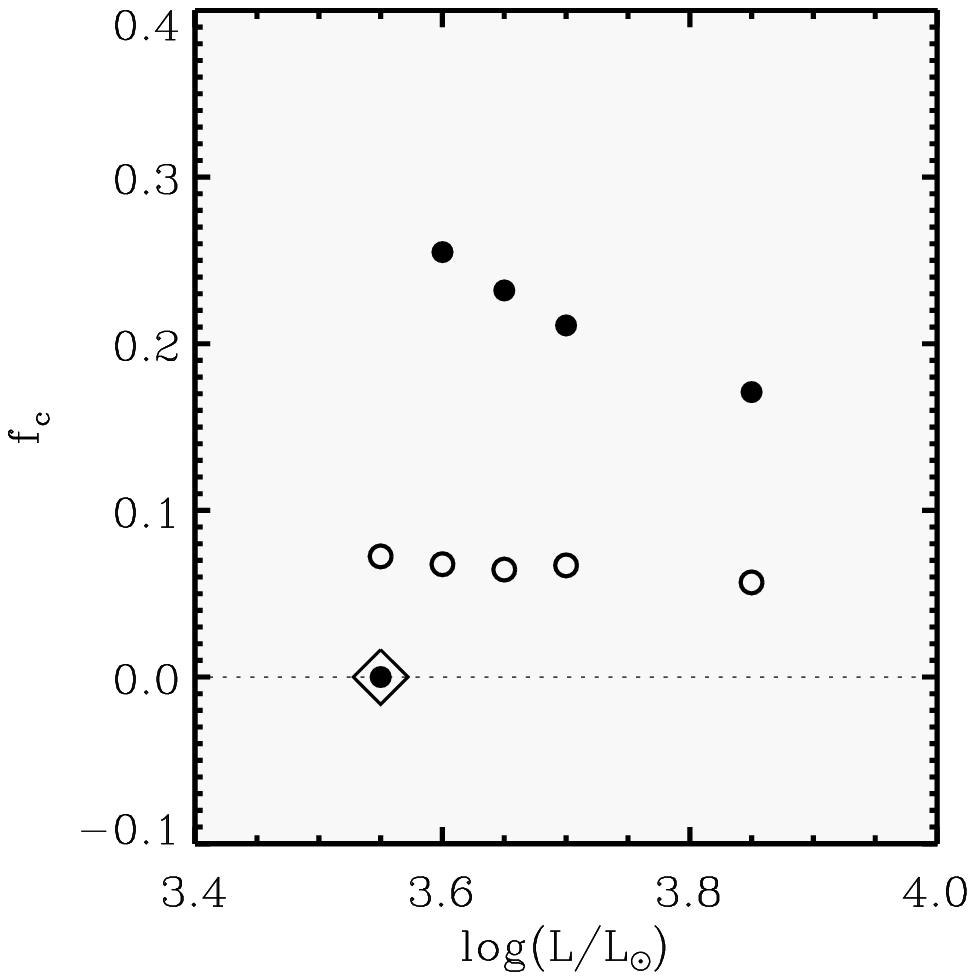}}
 \caption{Mass-loss rates (left panel), wind speeds (middle panel) and mean degree of dust condensation near a "threshold" as function of
 luminosity. All other stellar parameters are kept the same ($M_\star=1M_\odot$, $T_{\rm eff} = 2600$ K, $\log({\rm C}-{\rm O})= 8.80$). \label{thresholds}}
 \end{figure*}

\subsection{Thresholds for dust-driven winds}

In Paper~I we argued that for a realistic description of mass
loss it is also crucial to know in which parts of stellar parameter
space dust-driven mass loss cannot be sustained. Thresholds for
dust-driven outflows originate from the simple fact that there
exists a critical radiative-to-gravitational acceleration ratio
and thus critical values of
stellar parameters and element abundances for which the flux-mean
opacity equals $\kappa_{\rm crit}$ \cite{Dominik90,Mattsson07b}.
In Fig. \ref{thresholds} a threshold in luminosity is shown as an
example.
The transition regions in parameter space from windless models
to strong outflows are not adequately covered with the grid
spacing chosen in Paper I. To better resolve the transition region
and to quantify the effect of grain-size dependent opacities on the
threshold, we have computed a number of additional models close to
the expected mass-loss threshold taking smaller steps in
$\log(L_\star)$ while keeping all other parameters constant.
It is evident from the selected models that if the dust grains have
the most favorable size (in terms of achieving the highest possible
opacity, i.e., using optimized type A opacities), it will be
easier to sustain an outflow near a threshold. It is also clear that
the wind speed near a mass-loss threshold is significantly affected 
if the typical grain size is in the optimal range. In Fig.
\ref{thresholds} the wind speed is increased by approximately a
factor of two relative to the corresponding SPL models. The
mass-loss rate differs most at the lowest luminosity,
$\log(L_\star/L_\odot) = 3.55$, in which case grain-size effects
mean the difference between no outflow and a sustained dust-driven
wind. At the upper end of the tested luminosities, i.e.
$\log(L_\star/L_\odot) = 3.85$, the mass-loss rate is about 50\%
higher than in the SPL case, while it is almost unaffected for the
intermediate values of $L_\star$. The mean degree of dust
condensation $f_{\rm c}$ is roughly a factor of two lower
compared to the SPL models, like in the critical case models with
optimized type A opacities discussed above.

\begin{table*}
 \caption{\label{data_spa} Input parameters ($L_\star$, $T_{\rm eff}$, log(C-O), $P$) and the
 resulting mean mass-loss rate, mean velocity at the outer boundary and mean degree of dust condensation at the outer boundary, for a
 subset of models with $M_\star = 1 M_\odot$ and $\Delta u_{\rm p} = 4.0 \mbox{km s}^{-1}$ and the SPL used in the dust opacities.
 The dust-to-gas mass ratio $\rho_{\rm dust}/\rho_{\rm gas}$ is calculated from $f_{\rm c}$ as described in H\"ofner \& Dorfi (1997) and
 the quoted grain radius is defined as $a_{\rm gr}=a_{\rm mon}\,(K_1/K_0)$, where $K_0, K_1$ are the zeroth and first moment of the grain-size distrubution,
 respectively.}
 \center
 \begin{tabular}{cccccccccccccc}
 \hline
 \hline

Mod.   & $\log(L_\star)$ & $T_\mathrm{eff}$  & log(C-O) & $P$ & $\langle\dot{M}\rangle$ & $\langle u_{\rm out} \rangle$ & $\langle {\rm f_c} \rangle$ &
    $\langle {\rho_{\rm d}/\rho_{\rm g}} \rangle$ & $\langle a_{\rm gr}\rangle$\\[1mm]
  & [$L_{\odot}$] &  [$\mbox{K}$] & & [days] & [$M_\odot$ yr$^{-1}$] & [km  s$^{-1}$] & & & [cm]\\

 \hline
\\
 1 &    3.70   & 2400   &    8.20   &   295   &   -          &   -          &   -          &   -          &   -         \\
 2 &    3.70   & 2400   &    8.50   &   295   &   1.05E-06   &   3.80E+00   &   2.44E-01   &   6.77E-04   &   2.32E-05  \\
 3 &    3.85   & 2400   &    8.20   &   393   &   -          &   -          &   -          &   -          &   -         \\
 4 &    3.85   & 2400   &    8.50   &   393   &   3.15E-06   &   7.76E+00   &   2.13E-01   &   5.91E-04   &   2.32E-05  \\
 5 &    4.00   & 2400   &    8.20   &   524   &   2.42E-06   &   8.67E-01   &   2.54E-01   &   3.53E-04   &   8.30E-05  \\
 6 &    4.00   & 2400   &    8.50   &   524   &   8.50E-06   &   1.14E+01   &   2.15E-01   &   5.96E-04   &   2.48E-05  \\
 7 &    3.70   & 2600   &    8.50   &   295   &   7.60E-07   &   6.24E+00   &   2.11E-01   &   5.85E-04   &   ***       \\
 8 &    3.85   & 2600   &    8.20   &   393   &   -          &   -          &   -          &   -          &   -         \\
 9 &    3.85   & 2600   &    8.50   &   393   &   1.67E-06   &   6.24E+00   &   1.71E-01   &   4.74E-04   &   1.87E-05  \\
10 &    4.00   & 2600   &    8.20   &   524   &   1.22E-06   &   1.34E+00   &   2.06E-01   &   2.86E-04   &   5.52E-05  \\
11 &    4.00   & 2800   &    8.20   &   524   &   -          &   -          &   -          &   -          &   -         \\
12 &    3.70   & 3000   &    8.80   &   295   &   -          &   -          &   -          &   -          &   -         \\
\\
13 &    3.85   & 2400   &    8.80   &   393   &   5.40E-06   &   2.19E+01   &   2.85E-01   &   1.58E-03   &   1.15E-05  \\
14 &    3.85   & 2400   &    9.10   &   393   &   7.95E-06   &   4.15E+01   &   5.24E-01   &   5.79E-03   &   6.19E-06  \\
15 &    4.00   & 2400   &    8.80   &   524   &   1.23E-05   &   2.42E+01   &   3.36E-01   &   1.86E-03   &   1.26E-05  \\
16 &    4.00   & 2400   &    9.10   &   524   &   1.59E-05   &   4.43E+01   &   5.88E-01   &   6.49E-03   &   6.70E-06  \\
17 &    3.85   & 2600   &    8.80   &   393   &   4.04E-06   &   2.65E+01   &   3.22E-01   &   1.78E-03   &   1.24E-05  \\
18 &    3.85   & 2600   &    9.10   &   393   &   5.66E-06   &   4.65E+01   &   5.25E-01   &   5.80E-03   &   6.20E-06  \\
19 &    4.00   & 2600   &    8.80   &   524   &   8.43E-06   &   2.80E+01   &   3.25E-01   &   1.80E-03   &   1.18E-05  \\
20 &    4.00   & 2600   &    9.10   &   524   &   1.09E-05   &   4.55E+01   &   4.88E-01   &   5.39E-03   &   6.26E-06  \\
21 &    4.00   & 2800   &    8.80   &   524   &   5.76E-06   &   2.81E+01   &   2.94E-01   &   1.63E-03   &   1.33E-05  \\
22 &    4.00   & 2800   &    9.10   &   524   &   8.07E-06   &   4.96E+01   &   3.87E-01   &   4.27E-03   &   6.12E-06  \\
23 &    4.00   & 3000   &    8.80   &   524   &   3.85E-06   &   2.73E+01   &   2.86E-01   &   1.58E-03   &   1.21E-05  \\
24 &    4.00   & 3000   &    9.10   &   524   &   8.00E-06   &   4.76E+01   &   5.56E-01   &   6.14E-03   &   7.43E-06  \\
\\
 \hline
\\
 \end{tabular}
 \label{models}
 \end{table*}

\section{Conclusions}

In the first paper of this series we presented a large grid of
frequency-dependent dynamic models for atmospheres and winds of
C-type AGB stars, with the main purpose of providing a more
realistic description of dust-driven mass loss as input for stellar
evolution models. One of the underlying assumptions of the models,
i.e., that dust opacities can be described with the small-particle
limit (SPL) of the Mie theory, however, turned out to be questionable with
typical emerging grain sizes in a range where the SPL may severely
underestimate the actual grain opacities.

Introducing a generalized description of radiative cross sections
valid for arbitrary particle radii, we have explored grain size
effects on the wind properties of carbon stars. The time-dependent
description of grain growth used in our models readily gives
particle radii corresponding to various means of the grain size
distribution in every layer at every instant of time. To keep the
computational effort at a level that will allow the construction of
large model grids, we have used descriptions of the dust opacities
based on these local mean grain radii.

From the large number of models presented in Paper~I we selected two
samples, i.e., a group of models with strong, well-developed outflows
and another group close to thresholds for dust-driven winds in
stellar parameters space, referred to as "critical cases". For the
first group, which is presumably representative of most wind-forming
models in Paper~I, the effects of grain size on mass-loss rates and wind
velocities are small, whereas the critical models show more intense
mass loss and (in some cases significantly) higher wind velocities
when using the new, generalized description of dust opacities.
Models in both groups tend towards lower dust-to-gas ratios,
illustrating a self-regulating feedback between grain growth and the
increased opacity (and, consequently, higher radiative acceleration)
per dust mass. Therefore, in general, the "dust-loss rates" can be
expected to be lower than in Paper~I.

Extrapolating from the samples investigated here, it seems that the
mass loss rates given in Paper~I are reliable, within the limits of
current theoretical and observational uncertainties, except for
models close to thresholds for dust-driven outflows in stellar
parameter space where mass loss is probably underestimated. 

Given the results presented here, a full implementation of
grain-size-dependent opacities seems to be
important as future work. Furthermore, the actual sizes of dust
grains may be of great importance for theoretical spectra of dynamic
atmosphere models of carbon stars, which we plan to study in the
near future.

\begin{acknowledgements}
We thank B. Gustafsson for his comments on the original manuscript
draft. Both authors acknowledge support form the Swedish Research
Council ({\it Vetenskapsr{\aa}det}). The Dark Cosmology Centre is 
funded by the Danish National Research Foundation
\end{acknowledgements}

\begin{table*}
 \caption{\label{data_opt} Same as in Table \ref{data_spa}, but using the "$Q_{\rm rp}$-optimized" grain radius
 in the dust opacities.}
 \center
 \begin{tabular}{lccccccccccccc}
 \hline
 \hline

 Mod.   & $\log(L_\star)$ & $T_\mathrm{eff}$  & log(C-O) & $P$ & $\langle\dot{M}\rangle$ & $\langle u_{\rm out} \rangle$ & $\langle {\rm f_c} \rangle$ &
 $\langle {\rho_{\rm d}/\rho_{\rm g}} \rangle$ & $\langle a_{\rm gr}\rangle$\\[1mm]
 &[$L_{\odot}$] &  [$\mbox{K}$] & & [days] & [$M_\odot$ yr$^{-1}$] & [km  s$^{-1}$] & & & [cm]\\

 \hline
\\
 1 &   3.70   & 2400   & 8.20   &  295   &   4.42E-07   &   1.77E+00   &   9.76E-02   &   1.33E-04   &   5.05E-05  \\
 2 &   3.70   & 2400   & 8.50   &  295   &   1.82E-06   &   1.32E+01   &   7.58E-02   &   2.05E-04   &   1.80E-05  \\
 3 &   3.85   & 2400   & 8.20   &  393   &   1.89E-06   &   4.06E+00   &   7.24E-02   &   9.84E-05   &   4.84E-05  \\
 4 &   3.85   & 2400   & 8.50   &  393   &   4.43E-06   &   1.63E+01   &   8.86E-02   &   2.40E-04   &   1.95E-05  \\
 5 &   4.00   & 2400   & 8.20   &  524   &   5.40E-06   &   6.14E+00   &   5.90E-02   &   8.02E-05   &   4.51E-05  \\
 6 &   4.00   & 2400   & 8.50   &  524   &   8.68E-06   &   1.69E+01   &   7.00E-02   &   1.90E-04   &   1.84E-05  \\
 7 &   3.70   & 2600   & 8.50   &  295   &   9.27E-07   &   1.29E+01   &   6.29E-02   &   1.70E-04   &   1.66E-05  \\
 8 &   3.85   & 2600   & 8.20   &  393   &   5.90E-07   &   2.49E+00   &   6.50E-02   &   8.83E-05   &   4.24E-05  \\
 9 &   3.85   & 2600   & 8.50   &  393   &   2.44E-06   &   1.52E+01   &   5.77E-02   &   1.56E-04   &   1.57E-05  \\
10 &   4.00   & 2600   & 8.20   &  524   &   2.15E-06   &   4.43E+00   &   4.85E-02   &   6.59E-05   &   3.79E-05  \\
11 &   4.00   & 2800   & 8.20   &  524   &   -          &   -          &   -          &   -          &   -         \\
12 &   3.70   & 3000   & 8.80   &  295   &   2.69E-07   &   2.91E+01   &   6.32E-02   &   3.42E-04   &   6.52E-06  \\
\\
13 &   3.85   & 2400   & 8.80   &  393   &   5.15E-06   &   3.28E+01   &   1.19E-01   &   6.44E-04   &   8.20E-06  \\
14 &   3.85   & 2400   & 9.10   &  393   &   7.31E-06   &   5.48E+01   &   2.66E-01   &   2.87E-03   &   4.16E-06  \\
15 &   4.00   & 2400   & 8.80   &  524   &   1.03E-05   &   3.39E+01   &   1.23E-01   &   6.65E-04   &   8.12E-06  \\
16 &   4.00   & 2400   & 9.10   &  524   &   1.95E-05   &   5.39E+01   &   4.03E-01   &   4.35E-03   &   5.19E-06  \\
17 &   3.85   & 2600   & 8.80   &  393   &   3.32E-06   &   3.26E+01   &   1.05E-01   &   5.68E-04   &   8.15E-06  \\
18 &   3.85   & 2600   & 9.10   &  393   &   5.41E-06   &   5.54E+01   &   2.20E-01   &   2.37E-03   &   3.78E-06  \\
19 &   4.00   & 2600   & 8.80   &  524   &   6.54E-06   &   3.77E+01   &   1.14E-01   &   6.17E-04   &   8.21E-06  \\
20 &   4.00   & 2600   & 9.10   &  524   &   1.37E-05   &   5.42E+01   &   3.04E-01   &   3.28E-03   &   4.45E-06  \\
21 &   4.00   & 2800   & 8.80   &  524   &   5.04E-06   &   3.24E+01   &   1.05E-01   &   5.68E-04   &   8.40E-06  \\
22 &   4.00   & 2800   & 9.10   &  524   &   9.45E-06   &   5.87E+01   &   2.55E-01   &   2.75E-03   &   3.99E-06  \\
23 &   4.00   & 3000   & 8.80   &  524   &   3.98E-06   &   2.53E+01   &   8.70E-02   &   4.71E-04   &   8.15E-06  \\
24 &   4.00   & 3000   & 9.10   &  524   &   5.50E-06   &   6.89E+01   &   2.10E-01   &   2.27E-03   &   3.79E-06  \\
\\
 \hline
\\
 \end{tabular}
 \end{table*}

\begin{table*}
 \caption{\label{data_agr1} Same as in Table \ref{data_spa}, but using the first ($K_1$) mean grain radius in the
 dust opacities.}
 \center
 \begin{tabular}{lccccccccccccc}
 \hline
 \hline

 Mod.   & $\log(L_\star)$ & $T_\mathrm{eff}$  & log(C-O) & $P$ & $\langle\dot{M}\rangle$ & $\langle u_{\rm out} \rangle$ & $\langle {\rm f_c} \rangle$ &
 $\langle {\rho_{\rm d}/\rho_{\rm g}} \rangle$ & $\langle a_{\rm gr}\rangle$\\[1mm]
 &[$L_{\odot}$] &  [$\mbox{K}$] & & [days] & [$M_\odot$ yr$^{-1}$] & [km  s$^{-1}$] & & & [cm]\\

 \hline
\\
 1 &   3.70   & 2400   & 8.20   &  295   &   -          &   -          &   -          &   -          &   -         \\
 2 &   3.70   & 2400   & 8.50   &  295   &   1.65E-06   &   1.99E+01   &   1.35E-01   &   3.66E-04   &   2.60E-05  \\
 3 &   3.85   & 2400   & 8.20   &  393   &   2.23E-06   &   5.52E+00   &   8.66E-02   &   1.18E-04   &   4.57E-05  \\
 4 &   3.85   & 2400   & 8.50   &  393   &   3.85E-06   &   1.97E+01   &   1.44E-01   &   3.90E-04   &   2.57E-05  \\
 5 &   4.00   & 2400   & 8.20   &  524   &   6.46E-06   &   8.28E+00   &   7.06E-02   &   9.59E-05   &   3.88E-05  \\
 6 &   4.00   & 2400   & 8.50   &  524   &   9.03E-06   &   1.71E+01   &   1.27E-01   &   3.44E-04   &   2.25E-05  \\
 7 &   3.70   & 2600   & 8.50   &  295   &   6.45E-07   &   1.62E+01   &   8.68E-02   &   2.35E-04   &   1.99E-05  \\
 8 &   3.85   & 2600   & 8.20   &  393   &   -          &   -          &   -          &   -          &   -         \\
 9 &   3.85   & 2600   & 8.50   &  393   &   1.84E-06   &   1.86E+01   &   9.97E-02   &   2.70E-04   &   2.02E-05  \\
10 &   4.00   & 2600   & 8.20   &  524   &   1.49E-06   &   4.43E+00   &   5.00E-02   &   6.79E-05   &   4.14E-05  \\
11 &   4.00   & 2800   & 8.20   &  524   &   -          &   -          &   -          &   -          &   -         \\
12 &   3.70   & 3000   & 8.80   &  295   &   -          &   -          &   -          &   -          &   -         \\
\\
13 &   3.85   & 2400      & 8.80     & 393     & 6.41E-06     & 2.31E+01     & 2.61E-01     & 1.49E-03     & 1.27E-05  \\
14 &   3.85   & 2400      & 9.10     & 393     & 1.13E-05     & 4.28E+01     & 6.24E-01     & 7.09E-03     & 7.52E-06  \\
15 &   4.00   & 2400      & 8.80     & 524     & 1.66E-05     & 3.03E+01     & 1.23E-01     & 7.00E-04     & 1.56E-05  \\
16 &   4.00   & 2400      & 9.10     & 524     & 2.02E-05     & 4.61E+01     & 6.77E-01     & 7.69E-03     & 7.88E-06  \\
17 &   3.85   & 2600      & 8.80     & 393     & 4.15E-06     & 2.61E+01     & 2.64E-01     & 1.50E-03     & 1.24E-05  \\
18 &   3.85   & 2600      & 9.10     & 393     & 6.94E-06     & 4.64E+01     & 5.58E-01     & 6.34E-03     & 7.06E-06  \\
19 &   4.00   & 2600      & 8.80     & 524     & 9.22E-06     & 2.97E+01     & 3.04E-01     & 1.73E-03     & 1.33E-05  \\
20 &   4.00   & 2600      & 9.10     & 524     & 1.22E-05     & 5.14E+01     & 6.25E-01     & 7.10E-03     & 7.24E-06  \\
21 &   4.00   & 2800      & 8.80     & 524     & 6.32E-06     & 2.76E+01     & 2.46E-01     & 1.40E-03     & 1.22E-05  \\
22 &   4.00   & 2800      & 9.10     & 524     & 8.40E-06     & 5.28E+01     & 5.22E-01     & 5.93E-03     & 6.30E-06  \\
23 &   4.00   & 3000      & 8.80     & 524     & 3.87E-06     & 2.08E+01     & 2.00E-01     & 1.14E-03     & 1.09E-05  \\
24 &   4.00   & 3000      & 9.10     & 524     & 7.76E-06     & 4.51E+01     & 4.65E-01     & 5.28E-03     & 6.33E-06  \\
\\
 \hline
\\
 \end{tabular}
 \end{table*}

\begin{table*}
 \caption{\label{data_agr2} Same as in Table \ref{data_spa}, but using the second ($K_2$) mean grain radius in the
 dust opacities.}
 \center
 \begin{tabular}{lccccccccccccc}
 \hline
 \hline

 Mod.   & $\log(L_\star)$ & $T_\mathrm{eff}$  & log(C-O) & $P$ & $\langle\dot{M}\rangle$ & $\langle u_{\rm out} \rangle$ & $\langle {\rm f_c} \rangle$ &
 $\langle {\rho_{\rm d}/\rho_{\rm g}} \rangle$ & $\langle a_{\rm gr}\rangle$\\[1mm]
 & [$L_{\odot}$] &  [$\mbox{K}$] & & [days] & [$M_\odot$ yr$^{-1}$] & [km  s$^{-1}$] & & & [cm]\\

 \hline
\\
 1 &   3.70   & 2400   & 8.20   &  295   &   -          &   -          &   4.06E-01   &   5.52E-04   &   2.32E-04  \\
 2 &   3.70   & 2400   & 8.50   &  295   &   1.98E-06   &   1.71E+01   &   1.11E-01   &   3.01E-04   &   2.00E-05  \\
 3 &   3.85   & 2400   & 8.20   &  393   &   2.24E-06   &   4.97E+00   &   1.19E-01   &   1.62E-04   &   4.31E-05  \\
 4 &   3.85   & 2400   & 8.50   &  393   &   4.30E-06   &   1.72E+01   &   1.08E-01   &   2.93E-04   &   2.00E-05  \\
 5 &   4.00   & 2400   & 8.20   &  524   &   7.20E-06   &   8.12E+00   &   6.83E-02   &   9.28E-05   &   2.61E-05  \\
 6 &   4.00   & 2400   & 8.50   &  524   &   8.74E-06   &   1.59E+01   &   9.65E-02   &   2.62E-04   &   1.73E-05  \\
 7 &   3.70   & 2600   & 8.50   &  295   &   5.80E-07   &   1.45E+01   &   7.60E-02   &   2.06E-04   &   1.83E-05  \\
 8 &   3.85   & 2600   & 8.20   &  393   &   9.23E-07   &   4.97E+00   &   1.10E-01   &   1.49E-04   &   3.95E-05  \\
 9 &   3.85   & 2600   & 8.50   &  393   &   1.96E-06   &   1.87E+01   &   8.41E-02   &   2.28E-04   &   1.97E-05  \\
10 &   4.00   & 2600   & 8.20   &  524   &   3.59E-06   &   9.11E+00   &   8.13E-02   &   1.10E-04   &   3.60E-05  \\
11 &   4.00   & 2800   & 8.20   &  524   &   -          &   -          &   -          &   -          &   -         \\
12 &   3.70   & 3000   & 8.80   &  295   &   -          &   -          &   -          &   -          &   -         \\
\\
13 &   3.85   & 2400      & 8.80     & 393     & 5.89E-06     & 3.02E+01     & 2.99E-01     & 1.70E-03     & 1.33E-05  \\
14 &   3.85   & 2400      & 9.10     & 393     & 1.10E-05     & 4.40E+01     & 6.19E-01     & 7.03E-03     & 7.44E-06  \\
15 &   4.00   & 2400      & 8.80     & 524     & 1.47E-05     & 3.16E+01     & 3.78E-01     & 2.15E-03     & 1.45E-05  \\
16 &   4.00   & 2400      & 9.10     & 524     & 1.89E-05     & 4.71E+01     & 6.72E-01     & 7.63E-03     & 7.82E-06  \\
17 &   3.85   & 2600      & 8.80     & 393     & 3.77E-06     & 2.59E+01     & 2.23E-01     & 1.27E-03     & 1.16E-05  \\
18 &   3.85   & 2600      & 9.10     & 393     & 6.32E-06     & 4.82E+01     & 5.34E-01     & 6.07E-03     & 6.84E-06  \\
19 &   4.00   & 2600      & 8.80     & 524     & 8.44E-06     & 3.07E+01     & 2.60E-01     & 1.48E-03     & 1.23E-05  \\
20 &   4.00   & 2600      & 9.10     & 524     & 1.24E-05     & 5.14E+01     & 6.23E-01     & 7.08E-03     & 7.44E-06  \\
21 &   4.00   & 2800      & 8.80     & 524     & 5.89E-06     & 3.00E+01     & 2.32E-01     & 1.32E-03     & 1.18E-05  \\
22 &   4.00   & 2800      & 9.10     & 524     & 8.33E-06     & 5.34E+01     & 5.28E-01     & 6.00E-03     & 6.40E-06  \\
23 &   4.00   & 3000      & 8.80     & 524     & 3.28E-06     & 2.06E+01     & 1.71E-01     & 9.73E-04     & 1.04E-05  \\
24 &   4.00   & 3000      & 9.10     & 524     & 8.23E-06     & 4.58E+01     & 4.69E-01     & 5.33E-03     & 6.38E-06  \\
\\
 \hline
\\
 \end{tabular}
 \end{table*}

\begin{table*}
 \caption{\label{data_agr3} Same as in Table \ref{data_spa}, but using the third ($K_3$) mean grain radius in the
 dust opacities.}
 \center
 \begin{tabular}{lccccccccccccc}
 \hline
 \hline

 Mod.   & $\log(L_\star)$ & $T_\mathrm{eff}$  & log(C-O) & $P$ & $\langle\dot{M}\rangle$ & $\langle u_{\rm out} \rangle$ & $\langle {\rm f_c} \rangle$ &
 $\langle {\rho_{\rm d}/\rho_{\rm g}} \rangle$ & $\langle a_{\rm gr}\rangle$\\[1mm]
 &[$L_{\odot}$] &  [$\mbox{K}$] & & [days] & [$M_\odot$ yr$^{-1}$] & [km  s$^{-1}$] & & & [cm]\\

 \hline
\\
 1 &   3.70   & 2400   & 8.20   &  295   &   -          &   -          &   4.73E-01   &   6.43E-04   &   2.85E-04  \\
 2 &   3.70   & 2400   & 8.50   &  295   &   2.54E-06   &   1.57E+01   &   8.18E-02   &   2.22E-04   &   1.03E-05  \\
 3 &   3.85   & 2400   & 8.20   &  393   &   -          &   -          &   5.15E-01   &   7.00E-04   &   2.96E-04  \\
 4 &   3.85   & 2400   & 8.50   &  393   &   5.18E-06   &   1.73E+01   &   7.37E-02   &   2.00E-04   &   1.18E-05  \\
 5 &   4.00   & 2400   & 8.20   &  524   &   -          &   -          &   6.27E-01   &   8.52E-04   &   3.73E-04  \\
 6 &   4.00   & 2400   & 8.50   &  524   &   9.03E-06   &   2.19E+01   &   8.25E-02   &   2.24E-04   &   7.65E-06  \\
 7 &   3.70   & 2600   & 8.50   &  295   &   5.75E-07   &   1.43E+01   &   7.14E-02   &   1.94E-04   &   1.78E-05  \\
 8 &   3.85   & 2600   & 8.20   &  393   &   -          &   -          &   4.99E-01   &   6.78E-04   &   2.49E-04  \\
 9 &   3.85   & 2600   & 8.50   &  393   &   2.53E-06   &   1.74E+01   &   5.74E-02   &   1.56E-04   &   6.73E-06  \\
10 &   4.00   & 2600   & 8.20   &  524   &   1.91E-06   &   3.40E+00   &   1.86E-01   &   2.53E-04   &   2.01E-05  \\
11 &   4.00   & 2800   & 8.20   &  524   &   -          &   -          &   -          &   -          &   -         \\
12 &   3.70   & 3000   & 8.80   &  295   &   -          &   -          &   -          &   -          &   -         \\
\\
13 &   3.85   & 2400      & 8.80     & 393     & 5.84E-06     & 2.90E+01     & 2.48E-01     & 1.41E-03     & 1.22E-05  \\
14 &   3.85   & 2400      & 9.10     & 393     & 1.09E-05     & 4.30E+01     & 5.74E-01     & 6.52E-03     & 7.20E-06  \\
15 &   4.00   & 2400      & 8.80     & 524     & 1.44E-05     & 2.94E+01     & 2.93E-01     & 1.67E-03     & 1.24E-05  \\
16 &   4.00   & 2400      & 9.10     & 524     & 2.01E-05     & 4.71E+01     & 6.60E-01     & 7.50E-03     & 7.93E-06  \\
17 &   3.85   & 2600      & 8.80     & 393     & 3.61E-06     & 2.90E+01     & 2.71E-01     & 1.54E-03     & 1.12E-05  \\
18 &   3.85   & 2600      & 9.10     & 393     & 6.57E-06     & 4.76E+01     & 5.25E-01     & 5.96E-03     & 6.79E-06  \\
19 &   4.00   & 2600      & 8.80     & 524     & 8.68E-06     & 3.13E+01     & 2.40E-01     & 1.37E-03     & 1.18E-05  \\
20 &   4.00   & 2600      & 9.10     & 524     & 1.33E-05     & 5.06E+01     & 6.14E-01     & 6.97E-03     & 7.45E-06  \\
21 &   4.00   & 2800      & 8.80     & 524     & 5.31E-06     & 3.01E+01     & 1.85E-01     & 1.05E-03     & 1.04E-05  \\
22 &   4.00   & 2800      & 9.10     & 524     & 8.10E-06     & 5.54E+01     & 5.23E-01     & 5.94E-03     & 6.25E-06  \\
23 &   4.00   & 3000      & 8.80     & 524     & 3.58E-06     & 1.98E+01     & 1.44E-01     & 8.20E-04     & 9.75E-06  \\
24 &   4.00   & 3000      & 9.10     & 524     & 8.12E-06     & 4.91E+01     & 4.68E-01     & 5.32E-03     & 6.32E-06  \\
\\
 \hline
\\
 \end{tabular}
 \end{table*}

\bibliographystyle{aa}

\end{document}